\documentclass[reprint,twocolumn]{revtex4}
\usepackage{amsfonts}
\usepackage{amsmath}
\usepackage{amssymb}
\usepackage{charter}
\usepackage{graphicx}
\usepackage{float}
\usepackage{amsmath}
\usepackage{amssymb}
\usepackage{charter}
\usepackage{graphicx}
\usepackage{subfigure}
\usepackage{graphicx}
\usepackage{epstopdf}
\usepackage{color}
\usepackage{tocvsec2}
\usepackage{enumerate}
\usepackage{graphicx}
\usepackage{dcolumn}
\usepackage{bm}

\begin{document}

\title{Enhanced Phase Estimation via Photon-Added Two-Mode Squeezed States
and Kerr Nonlinearity}
\author{Zekun Zhao$^{1}$}
\author{Qingqian Kang$^{1,2}$}
\author{Teng Zhao$^{1}$}
\author{Cunjin Liu$^{1}$}
\author{Xin Su$^{1}$}
\author{Liyun Hu$^{1,3}$}
\thanks{Corresponding author: hlyun@jxnu.edu.cn}
\affiliation{$^{1}$\textit{Center for Quantum Science and Technology, Jiangxi Normal
University, Nanchang 330022, China}\\
$^{2}$\textit{Department of Physics, Jiangxi Normal University Science and
Technology College, Nanchang 330022, China}\\
$^{3}$\textit{Institute for Military-Civilian Integration of Jiangxi
Province, Nanchang 330200, China} }

\begin{abstract}
Quantum metrology employs quantum resources to achieve measurement precision
beyond classical limits. This work investigates a Mach--Zehnder
interferometer incorporating a Kerr nonlinear phase shifter, with
photon-added two-mode squeezed coherent states generated via four-wave
mixing as input. We demonstrate that increasing both the photon-addition
order and the input resource strength systematically enhances phase
sensitivity, quantum Fisher information, and the corresponding quantum
Cram\'er--Rao bound. The proposed system not only surpasses the standard
quantum limit but also approaches or exceeds the Heisenberg limit for linear
phase shifts, while Kerr nonlinearity enables surpassing the
super-Heisenberg limit. Furthermore, the scheme exhibits enhanced robustness
against photon loss, providing a promising pathway toward practical
high-precision quantum metrology applications.

\end{abstract}

\maketitle

\section{Introduction}

Precision measurement constitutes the cornerstone of scientific and
technological advancement, with its fundamental limits dictating our ability
to probe and manipulate physical phenomena. From traditional mechanical
manufacturing to modern lithography, and from gravitational-wave detection
to biomedical sensing, progress consistently relies on advancing measurement
precision. Within classical frameworks, measurement accuracy is
fundamentally constrained by the standard quantum limit (SQL), where
measurement uncertainty scales as $1/\sqrt{\bar{N}}$ with the average photon
number $\bar{N}$ inside the interferometer \cite{1,2,3}.

Quantum mechanics offers transformative strategies to overcome these
limitations. Quantum metrology exploits distinct quantum
resources---including superposition, entanglement, and nonclassical
correlations---to surpass the SQL and achieve ultra-high precision
approaching or exceeding the Heisenberg limit (HL), which scales as $1/\bar{N%
}$ \cite{4,5,6,7,8,9,10,11}. This field carries profound physical
implications and substantial potential for applications ranging from
gravitational-wave detection and atomic clocks to magnetometry,
super-resolution imaging, and measurements of fundamental physical constants.

A primary strategy for quantum enhancement involves injecting nonclassical
states into interferometric devices rather than conventional coherent
states. Squeezed states and entangled states generated via parametric
amplification processes have proven effective in surpassing the SQL.
However, preparing ideal entangled states (e.g., NOON states) remains
challenging due to their extreme susceptibility to environmental noise and
photon loss, which significantly limits practical implementation \cite%
{11,12,13,14,15,16}.

Recent research has extensively explored non-Gaussian operations,
particularly photon addition and subtraction, which can substantially
enhance the nonclassical properties of quantum states and potentially
improve measurement robustness. Among these, photon-addition operations
offer distinct advantages in quantum-state preparation and in enhancing both
nonclassical characteristics and measurement precision \cite%
{17,18,19,20,21,22,23,24,25,26,27,28,29,30,31,32}. Concurrently,
interferometer architectures have been continuously refined. For instance,
incorporating Kerr nonlinear media to replace conventional linear phase
shifters can theoretically achieve accuracies beyond the sub-HL ($1/\bar{N}%
^{3/2}$), potentially reaching the super-Heisenberg limit (SHL), i.e., $1/%
\bar{N}^{2}$ \cite{33,34,35,36,37,38,39,40,41}.

Building upon this foundation, this work develops a quantum precision
measurement scheme that combines high precision with enhanced robustness. We
generate two-mode squeezed coherent states (TMSCS) via four-wave mixing
(FWM) and apply photon-addition operations to create photon-added TMSCS
(PA-TMSCS) as the interferometer input resource. Furthermore, we replace the
linear phase shifter in a conventional Mach--Zehnder interferometer (MZI)
with a Kerr nonlinear phase shifter, constructing a Kerr nonlinear MZI
(KMZI). This study systematically analyzes the scheme's performance under
both ideal conditions and in the presence of photon loss. By examining phase
sensitivity via intensity-difference detection and quantum Fisher
information (QFI), and comparing linear and nonlinear phase-shift scenarios,
we verify the synergistic advantages of photon-addition operations and Kerr
nonlinear effects in enhancing measurement accuracy and system resilience.
This work provides valuable theoretical insights for developing practical
quantum-enhanced measurement technologies.

The paper is organized as follows: Section~II introduces the phase
estimation model based on a KMZI with PA-TMSCS input. Section~III
investigates the phase sensitivity of intensity-difference detection for
both linear ($k=1$) and Kerr nonlinear ($k=2$) phase shifts, including the
effects of photon loss. Section~IV examines the QFI for these cases under
ideal and lossy conditions. Conclusions are presented in the final section.

\section{Phase Estimation Model Based on a KMZI with PA-TMSCS Input}

\begin{figure}[t]
\centering
\includegraphics[width=0.83\columnwidth]{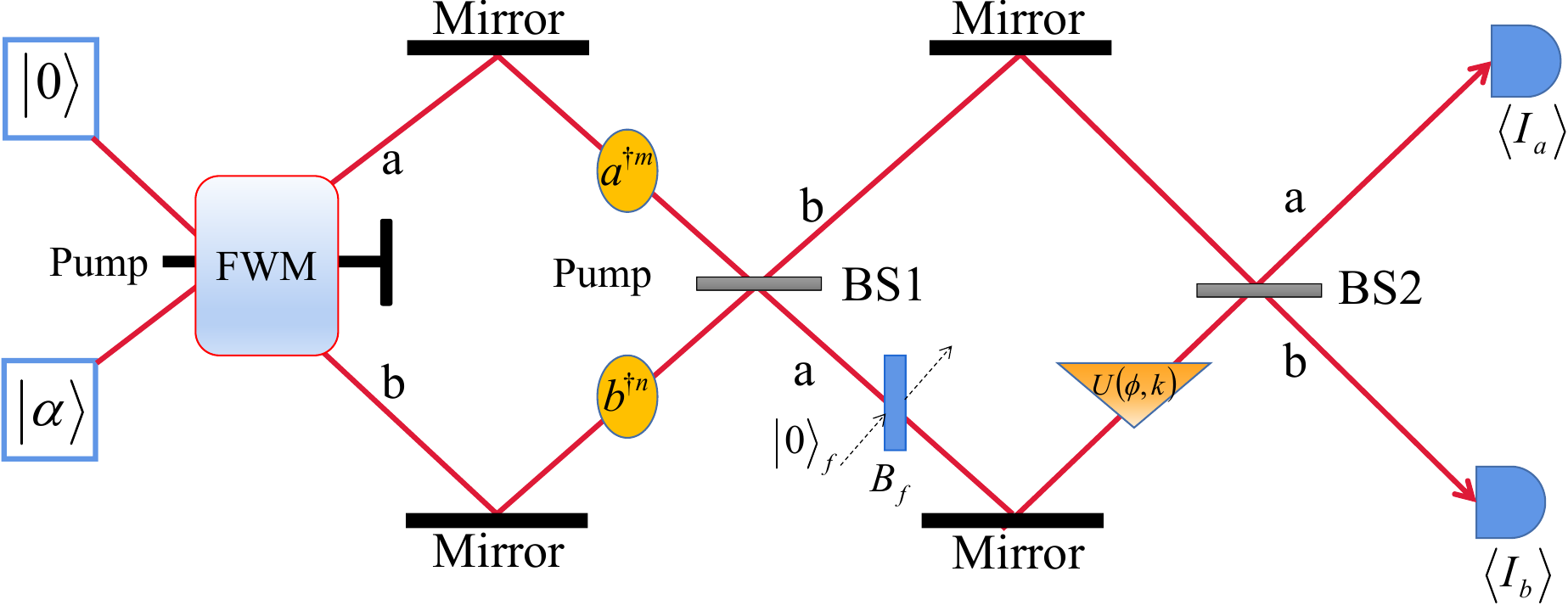}
\caption{Schematic of phase estimation using PA-TMSCS inputs in a KMZI.
Photon-addition operations are applied to both the $a$-mode and $b$-mode of
the TMSCS. Photon loss inside the KMZI is simulated by a virtual beam
splitter, and intensity-difference detection is implemented at the output
port.}
\label{Fig1}
\end{figure}

This section presents the phase estimation model employing PA-TMSCS inputs
in a KMZI. As illustrated in Fig.~\ref{Fig1}, TMSCS can be generated by
applying an FWM process induced by a pump field to coherent states. The FWM
process is equivalent to a two-mode squeezing operation described by the
operator $S_{2}(r) = \exp[\xi^{\ast} a b - \xi a^{\dagger} b^{\dagger}]$,
where $\xi = r e^{i\phi_{1}}$, with $r$ and $\phi_{1}$ representing the
squeezing parameter and phase, respectively. To simplify experimental
procedures and avoid complicated phase-matching conditions, only the $a$%
-mode input is prepared as a coherent state $\lvert \alpha \rangle_{a}$,
while the $b$-mode input remains in the vacuum state $\lvert 0\rangle_{b}$.
The coherent state satisfies $a \lvert \alpha \rangle_{a} = \alpha \lvert
\alpha \rangle_{a}$, where $\alpha = \lvert \alpha \rvert e^{i\theta}$ ($%
\lvert \alpha \rvert$ and $\theta$ denote amplitude and phase,
respectively). For simplicity and to satisfy phase-matching conditions, we
set $\theta = 0$ ($\alpha = \lvert \alpha \rvert$) and $\phi_{1} = \pi$,
which proves advantageous for enhanced phase estimation \cite{24,42}.

To further optimize the input states for improved measurement accuracy, we
implement non-Gaussian photon-addition operations on the TMSCS. The
resulting PA-TMSCS serves as the input state and can be expressed as
\begin{equation}
\lvert \text{in} \rangle = \frac{1}{\sqrt{N_{m,n}}} a^{\dagger m} b^{\dagger
n} S_{2}(r) \lvert \alpha \rangle_{a} \lvert 0 \rangle_{b},
\label{eq:in_state}
\end{equation}
where $a^{\dagger m}$ and $b^{\dagger n}$ represent photon-addition
operations on the $a$-mode and $b$-mode, respectively, and $N_{m,n}$ is the
normalization coefficient. To facilitate calculations, the photon-addition
operators in Eq.~(\ref{eq:in_state}) can be converted into
partial-derivative form:
\begin{eqnarray}
\lvert \text{in} \rangle &=& \frac{1}{\sqrt{N_{m,n}}} \frac{\partial^{m+n}}{%
\partial t_{1}^{m} \partial \tau_{1}^{n}} \exp[a^{\dagger} t_{1}]  \notag \\
&& \times \exp[b^{\dagger} \tau_{1}] S_{2}(r) \lvert \alpha \rangle_{a}
\lvert 0 \rangle_{b} \Big|_{t_{1}=\tau_{1}=0}.  \label{eq:in_state_deriv}
\end{eqnarray}
Based on this representation and quantum-state normalization, the expression
for $N_{m,n}$ can be derived through relevant operator transformation
relations:
\begin{equation}
\begin{aligned} N_{m,n} = \Gamma_{0} \Bigl\{ & \exp\bigl[((t_{1}t_{2} +
\tau_{1}\tau_{2}) \cosh r \\ & + (t_{1}\tau_{1} + t_{2}\tau_{2}) \sinh r)
\cosh r\bigr] \\ & \times \exp\bigl[((t_{1}+t_{2}) \cosh r \\ & +
(\tau_{1}+\tau_{2}) \sinh r) \alpha \bigr] \Bigr\}, \end{aligned}
\label{eq:norm_coeff}
\end{equation}
where
\begin{equation}
\Gamma_{0}\{\cdot\} = \frac{\partial^{2m+2n}}{\partial t_{1}^{m} \partial
\tau_{1}^{n} \partial t_{2}^{m} \partial \tau_{2}^{n}} \{\cdot\} \Big|%
_{t_{1}=\tau_{1}=t_{2}=\tau_{2}=0}.  \label{eq:Gamma0}
\end{equation}

Conventional MZIs constitute linear measuring instruments comprising two
50:50 beam splitters (BS1 and BS2) and a linear phase shifter. Following
Yurke \emph{et al.} \cite{43}, using angular momentum operators in the
Schwinger representation, the equivalent operators for BS1 and BS2 are $%
B_{1} = \exp[-i\pi (a^{\dagger}b + a b^{\dagger})/4]$ and $B_{2} = \exp[i\pi
(a^{\dagger}b + a b^{\dagger})/4]$, respectively. These operators satisfy
the transformation relations:
\begin{equation}
B_{1}^{\dagger} \binom{a}{b} B_{1} = \frac{\sqrt{2}}{2}
\begin{pmatrix}
1 & -i \\
-i & 1%
\end{pmatrix}
\binom{a}{b},  \label{eq:BS1_trans}
\end{equation}
and
\begin{equation}
B_{2}^{\dagger} \binom{a}{b} B_{2} = \frac{\sqrt{2}}{2}
\begin{pmatrix}
1 & i \\
i & 1%
\end{pmatrix}
\binom{a}{b}.  \label{eq:BS2_trans}
\end{equation}

For our investigation, we replace the linear phase shifter in the
conventional MZI with a Kerr nonlinear phase shifter to construct an
improved KMZI. The phase shifter's equivalent operator is $U(\phi,k) = \exp[%
i \phi (a^{\dagger}a)^{k}]$, where $k=1$ corresponds to a linear phase shift
and $k=2$ to a Kerr nonlinear phase shift. Here, $\phi$ represents the phase
shift to be measured. The transformation relations for linear and Kerr
nonlinear phase shifts are respectively given by:
\begin{equation}
U^{\dagger}(\phi,1) a U(\phi,1) = e^{i\phi} a,  \label{eq:linear_trans}
\end{equation}
and
\begin{equation}
U^{\dagger}(\phi,2) a U(\phi,2) = e^{i\phi} e^{i\phi (2 a^{\dagger}a)} a.
\label{eq:kerr_trans}
\end{equation}
The transformation relation for $U(\phi,2)$ in Eq.~(\ref{eq:kerr_trans}) can
be derived using methods from Ref.~\cite{40} based on Fock-state
representation, with detailed steps provided in Appendix~A.

During actual measurements, photon losses inevitably occur. In our model
(Fig.~\ref{Fig1}), we simulate photon losses in the KMZI's internal $a$-mode
using a virtual beam splitter placed between BS1 and the phase shifter. The
corresponding transformation is:
\begin{equation}
B_{f}^{\dagger}
\begin{pmatrix}
a \\
a_{f}%
\end{pmatrix}
B_{f} =
\begin{pmatrix}
\sqrt{1-l} & \sqrt{l} \\
-\sqrt{l} & \sqrt{1-l}%
\end{pmatrix}
\begin{pmatrix}
a \\
a_{f}%
\end{pmatrix}%
,  \label{eq:loss_trans}
\end{equation}
where $a_{f}$ is the photon annihilation operator for the auxiliary mode $f$
containing vacuum noise $\lvert 0 \rangle_{f}$, while $B_{f}$ and $l$
represent the equivalent operator and reflectivity of the virtual beam
splitter, respectively. The parameter $l$ corresponds to the photon loss
rate, satisfying $0 \le l \le 1$. Higher $l$ values indicate increased loss,
with $l=0$ representing the lossless case and $l=1$ corresponding to
complete absorption.

Within this KMZI framework, intensity-difference detection is implemented by
placing photon counters at the output ports. Since intensity-difference
detection is relatively straightforward to implement experimentally, it
finds widespread use for phase sensitivity measurements in MZIs.

\section{Phase Sensitivity of Intensity-Difference Detection}

The phase sensitivity for intensity-difference detection can be expressed
using the error propagation formula:
\begin{equation}
\Delta \phi = \frac{\sqrt{\langle I_{D}^{2} \rangle - \langle I_{D}
\rangle^{2}}}{\bigl\lvert \frac{\partial \langle I_{D} \rangle}{\partial \phi%
} \bigr\rvert},  \label{eq:phase_sensitivity}
\end{equation}
where $I_{D} = a^{\dagger}a - b^{\dagger}b$, and $\langle \cdot \rangle =
\langle \text{out} \rvert \cdot \lvert \text{out} \rangle$ denotes the
expectation value with respect to the output state. From the phase
estimation model description, the output state can be expressed as:
\begin{equation}
\lvert \text{out} \rangle = B_{2} U(\phi,k) B_{f} B_{1} \lvert \text{in}
\rangle \lvert 0 \rangle_{f}.  \label{eq:out_state}
\end{equation}
We focus on phase sensitivity for both $k=1$ and $k=2$ cases. For brevity,
explicit expressions for phase sensitivity are omitted here. Substituting
the expectation values $\langle I_{D} \rangle$ and $\langle I_{D}^{2}
\rangle $ into Eq.~(\ref{eq:phase_sensitivity}) yields the phase
sensitivity. The specific expressions for $\langle I_{D} \rangle$ and $%
\langle I_{D}^{2} \rangle$ under $k=1$ and $k=2$ conditions are provided in
Appendices~B and C, respectively.

\begin{figure*}[t]
\centering
\includegraphics[width=0.49\textwidth]{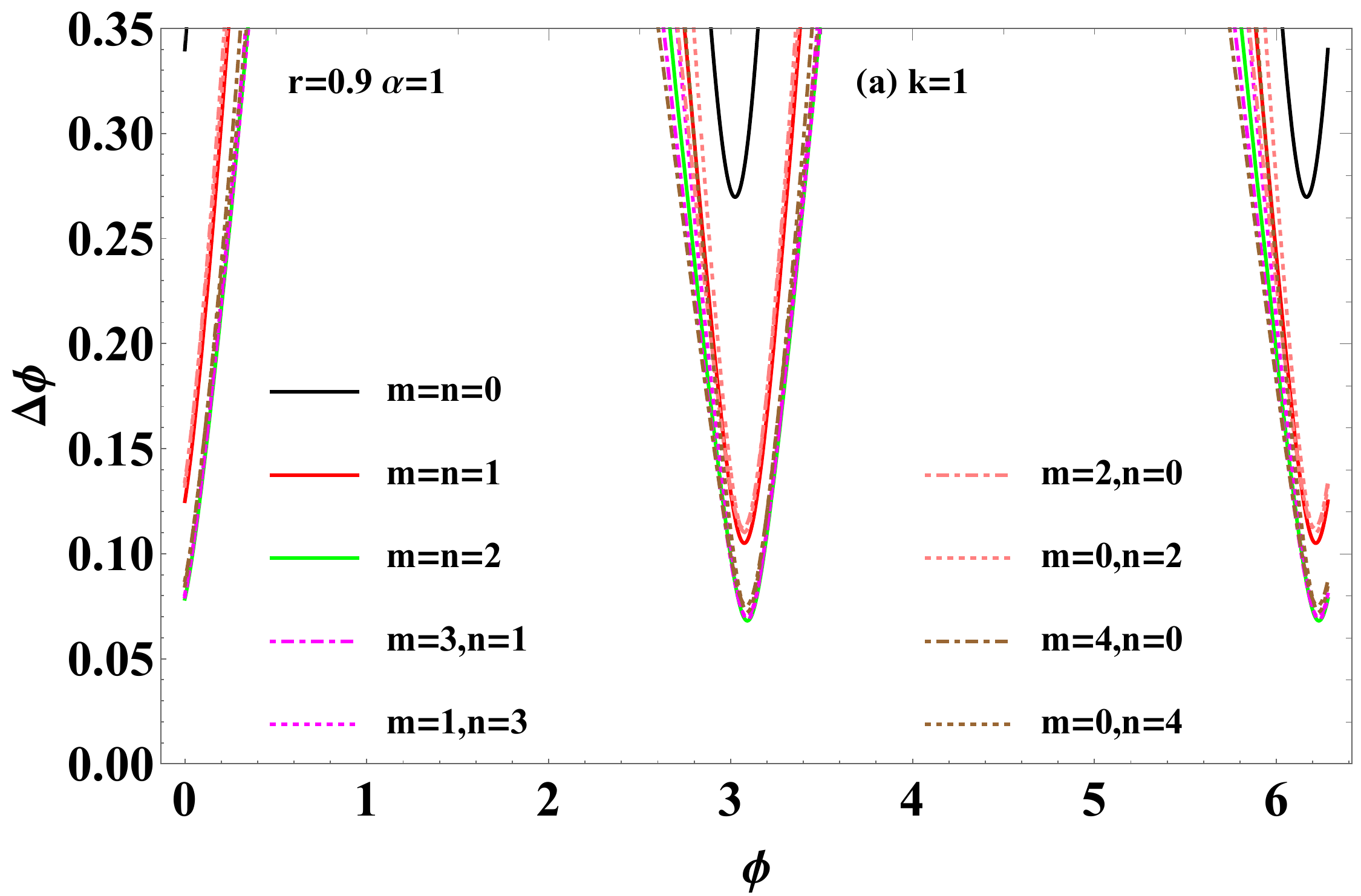} %
\includegraphics[width=0.49\textwidth]{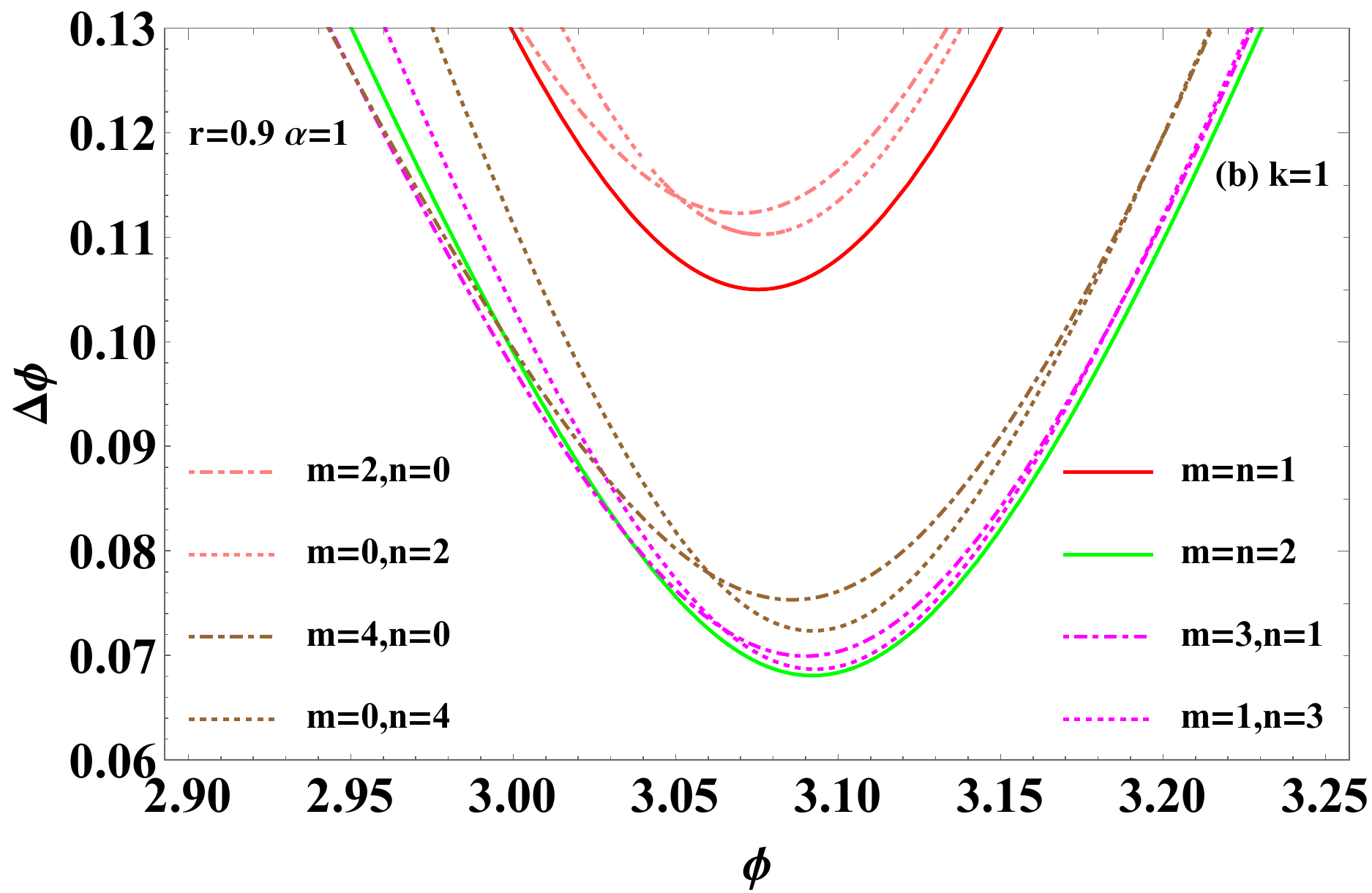} \newline
\includegraphics[width=0.49\textwidth]{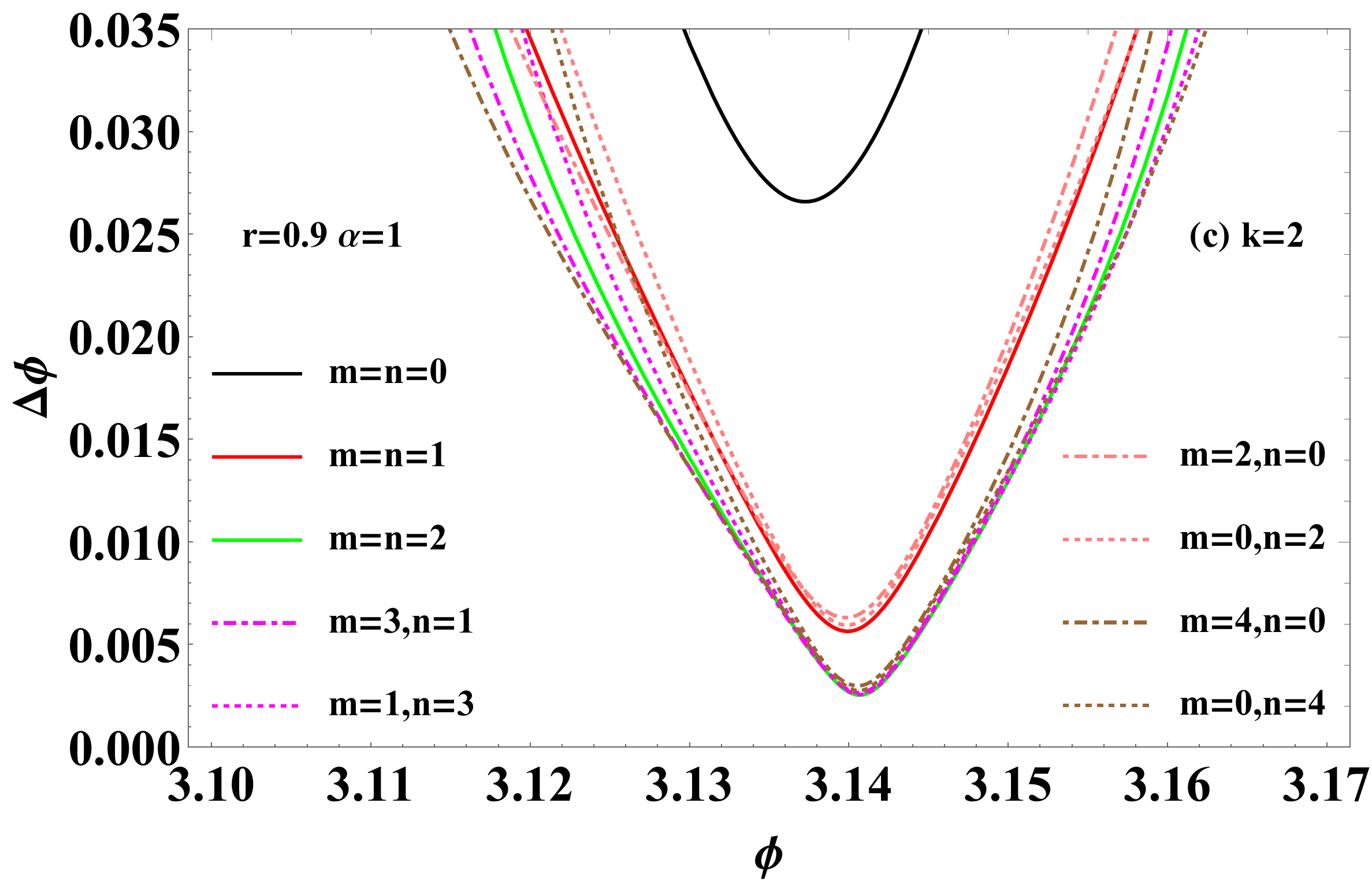} %
\includegraphics[width=0.49\textwidth]{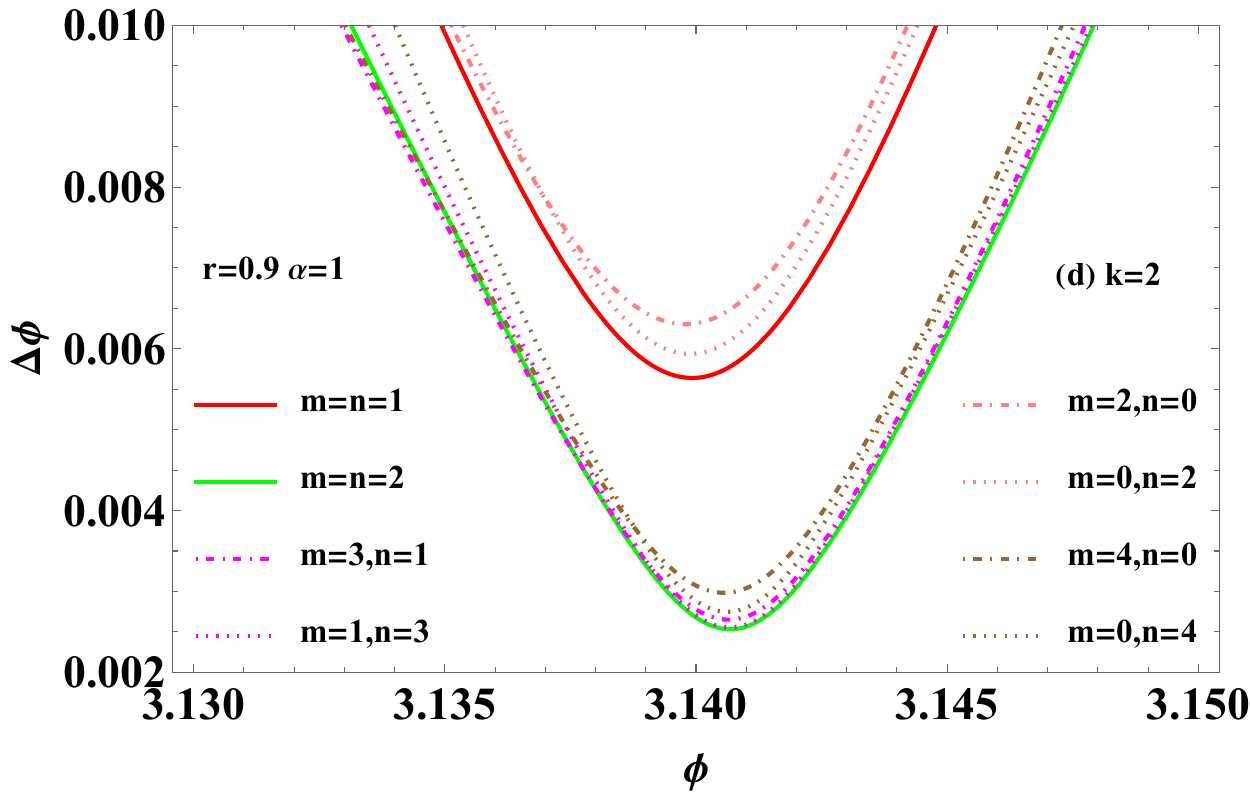}
\caption{Phase sensitivity $\Delta\protect\phi$ versus phase shift $\protect%
\phi$ for squeezing parameter $r=0.9$ and coherent amplitude $\protect\alpha%
=1$. (a) and (b): linear phase shift ($k=1$); (c) and (d): Kerr nonlinear
phase shift ($k=2$). The black solid line corresponds to $m=n=0$ (no photon
addition).}
\label{Fig2}
\end{figure*}

Fig.~\ref{Fig2} illustrates the variation of phase sensitivity $\Delta\phi$
with phase shift $\phi$ for $k=1$ and $k=2$ cases, where Figs.~\ref{Fig2}(a)
and (b) depict linear phase shift ($k=1$), while Figs.~\ref{Fig2}(c) and (d)
represent Kerr nonlinear phase shift ($k=2$). According to our computational
expressions and the behavior of $\Delta\phi$ in Fig.~\ref{Fig2}(a), the
variation period is $\pi$. Comparing Figs.~\ref{Fig2}(a) and (c), curves
incorporating photon-addition operations show significant improvement over
the black solid line ($m=n=0$). Thus, for schemes employing PA-TMSCS input
states, photon addition effectively enhances phase sensitivity. Notably, the
phase sensitivity variation range in Fig.~\ref{Fig2}(c) is substantially
reduced compared to Fig.~\ref{Fig2}(a) due to the Kerr nonlinear effect at $%
k=2$ (hence Fig.~\ref{Fig2}(c) is not plotted over a full period),
suggesting that Kerr nonlinear phase shifting can significantly improve
phase sensitivity within certain phase shift ranges.

To further investigate effective photon-addition schemes, we analyze phase
sensitivity characteristics under narrowed plotting ranges, as shown in
Figs.~\ref{Fig2}(b) and (d). Phase sensitivity improves with increasing
photon-addition order $m+n$, and near the optimal phase sensitivity point,
symmetric photon addition ($m=n$) yields the best performance for a given $%
m+n$. This indicates that performing symmetric photon-addition operations in
both $a$- and $b$-modes is more conducive to enhancing measurement accuracy.
Consequently, we focus on the $m=n$ scenario for photon-addition operations
throughout this paper.

\begin{figure}[t]
\centering
\includegraphics[width=0.8\columnwidth]{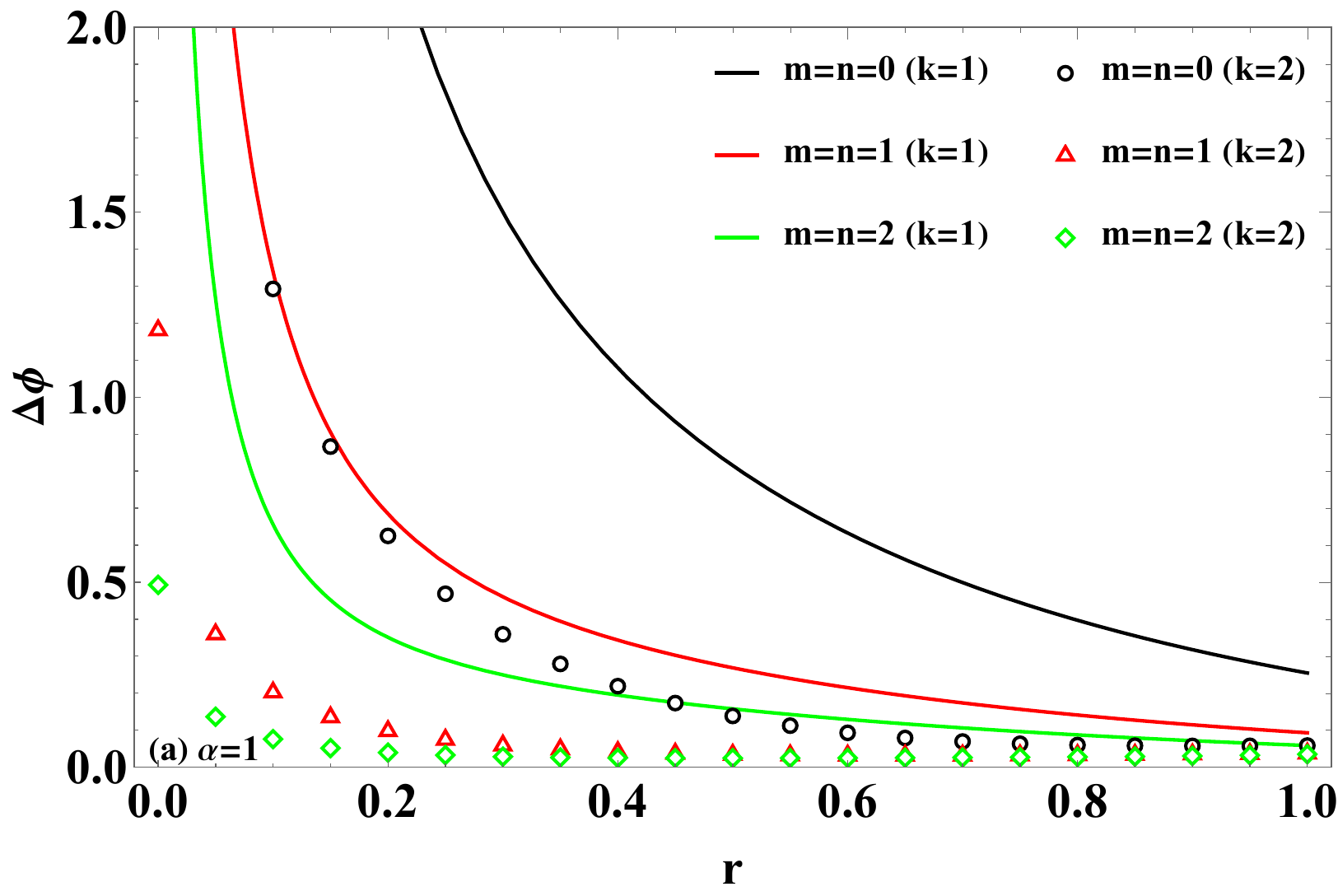} \newline
\includegraphics[width=0.8\columnwidth]{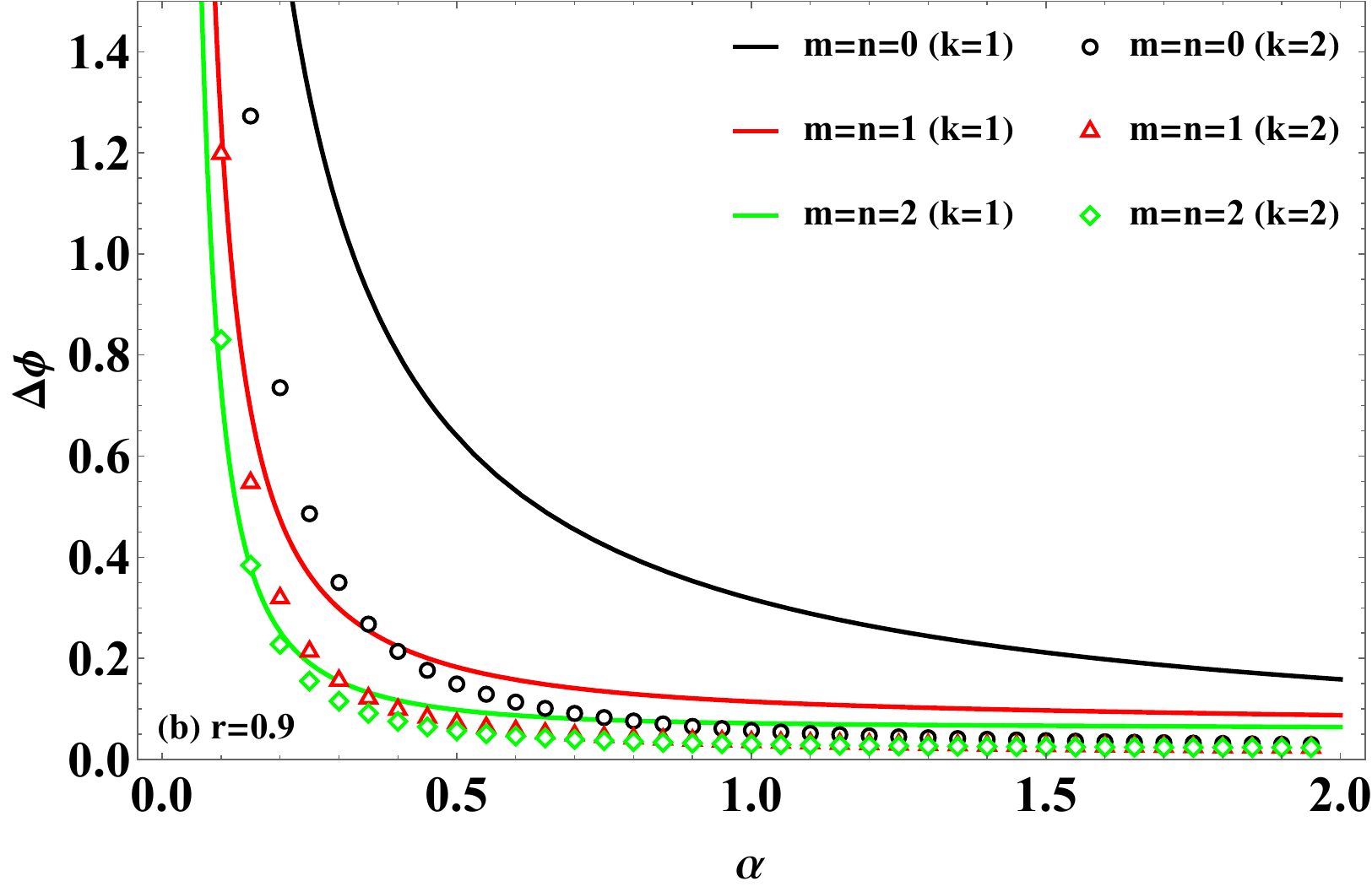}
\caption{For fixed phase shift $\protect\phi=3.12$: (a) phase sensitivity $%
\Delta\protect\phi$ versus squeezing parameter $r$ with coherent amplitude $%
\protect\alpha=1$; (b) $\Delta\protect\phi$ versus $\protect\alpha$ with $%
r=0.9$. Solid (dashed) lines correspond to $k=2$ ($k=1$).}
\label{Fig3}
\end{figure}

Fig.~\ref{Fig3} demonstrates that phase sensitivity $\Delta\phi$ improves
with increasing squeezing parameter $r$, coherent amplitude $\alpha$, and
photon-addition order. Additionally, comparing solid and dashed lines
clearly shows that Kerr nonlinear phase shift provides improved phase
sensitivity compared to linear phase shift.

In practical scenarios, photon loss constitutes an inevitable factor. We
subsequently analyze its impact on phase sensitivity and compare achievable
measurement precision limits. The measurement precision limit can be defined
through the average photon number inside the interferometer $\bar{N}$,
expressed as:
\begin{eqnarray}
\bar{N} &=& {}_{\text{int}}\langle \psi \rvert a^{\dagger}a + b^{\dagger}b
\lvert \psi \rangle_{\text{int}}  \notag \\
&=& D_{1}(1,0,1,0) + D_{1}(0,1,0,1),  \label{eq:avg_photon}
\end{eqnarray}
where $\lvert \psi \rangle_{\text{int}} = U(\phi,k) B_{f} B_{1} \lvert \text{%
in} \rangle \lvert 0 \rangle_{f}$, and specific expressions for $%
D_{1}(1,0,1,0)$ and $D_{1}(0,1,0,1)$ are given in Eq.~(B5) of Appendix~B.

\begin{figure}[t]
\centering
\includegraphics[width=0.8\columnwidth]{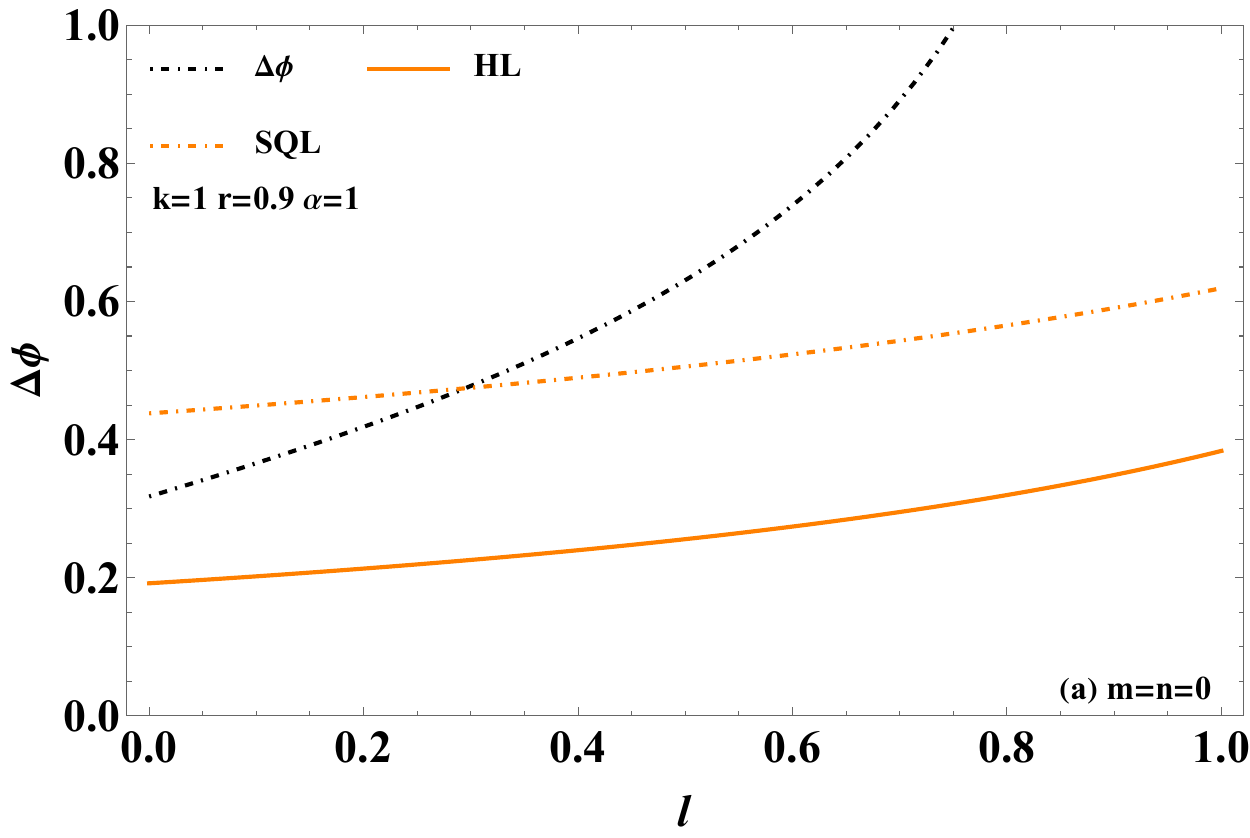} \newline
\includegraphics[width=0.8\columnwidth]{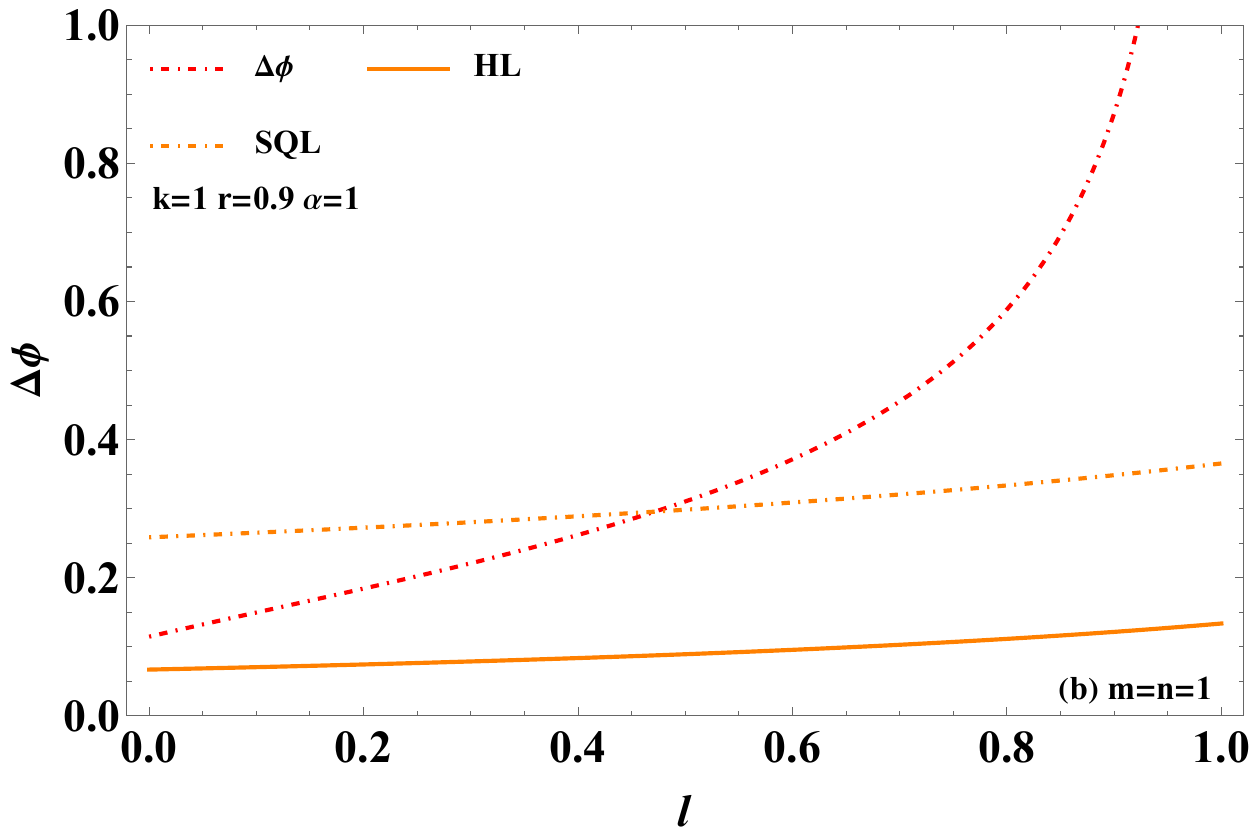} \newline
\includegraphics[width=0.8\columnwidth]{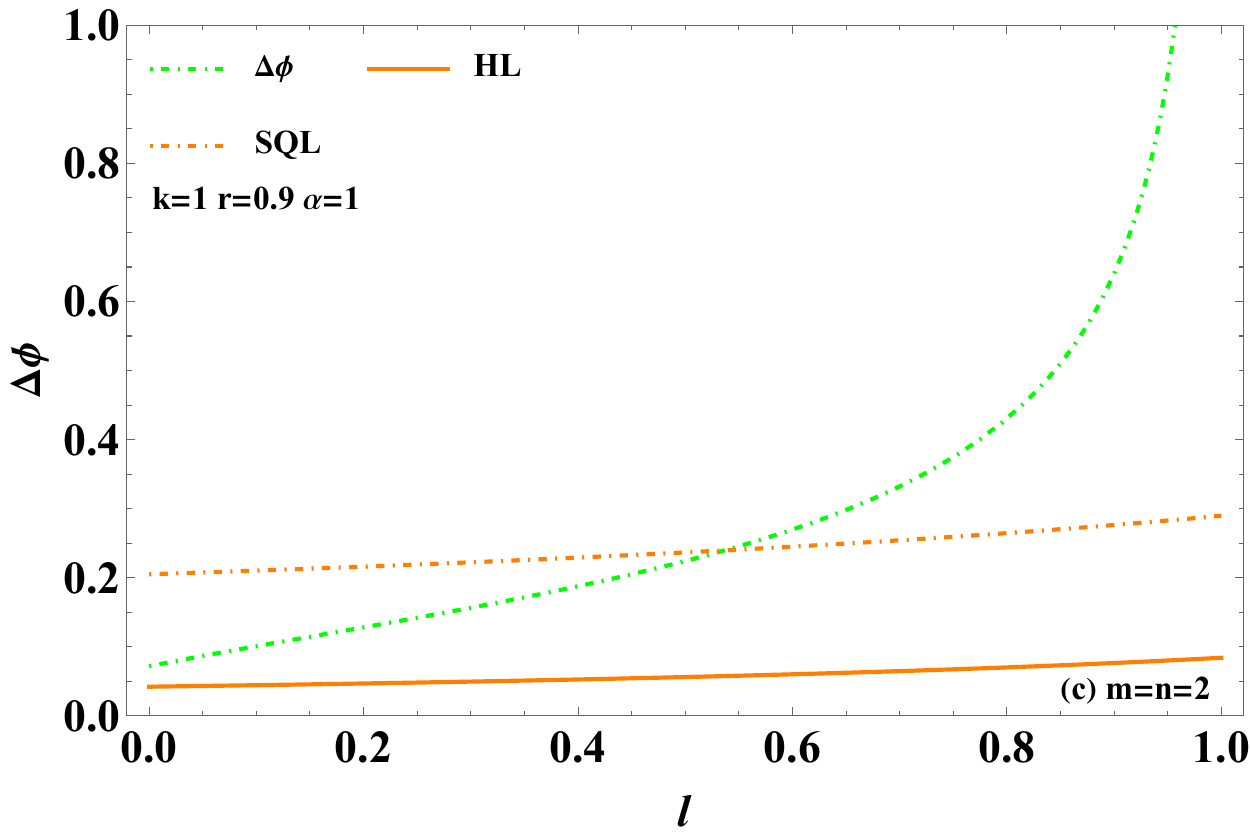}
\caption{Linear phase shift case ($k=1$): phase sensitivity $\Delta\protect%
\phi$ versus loss rate $l$ for (a) $m=n=0$, (b) $m=n=1$, (c) $m=n=2$, with
fixed phase shift $\protect\phi=3.12$, squeezing parameter $r=0.9$, and
coherent amplitude $\protect\alpha=1$. SQL and HL are shown for comparison.}
\label{Fig4}
\end{figure}

\begin{figure}[t]
\centering
\includegraphics[width=0.8\columnwidth]{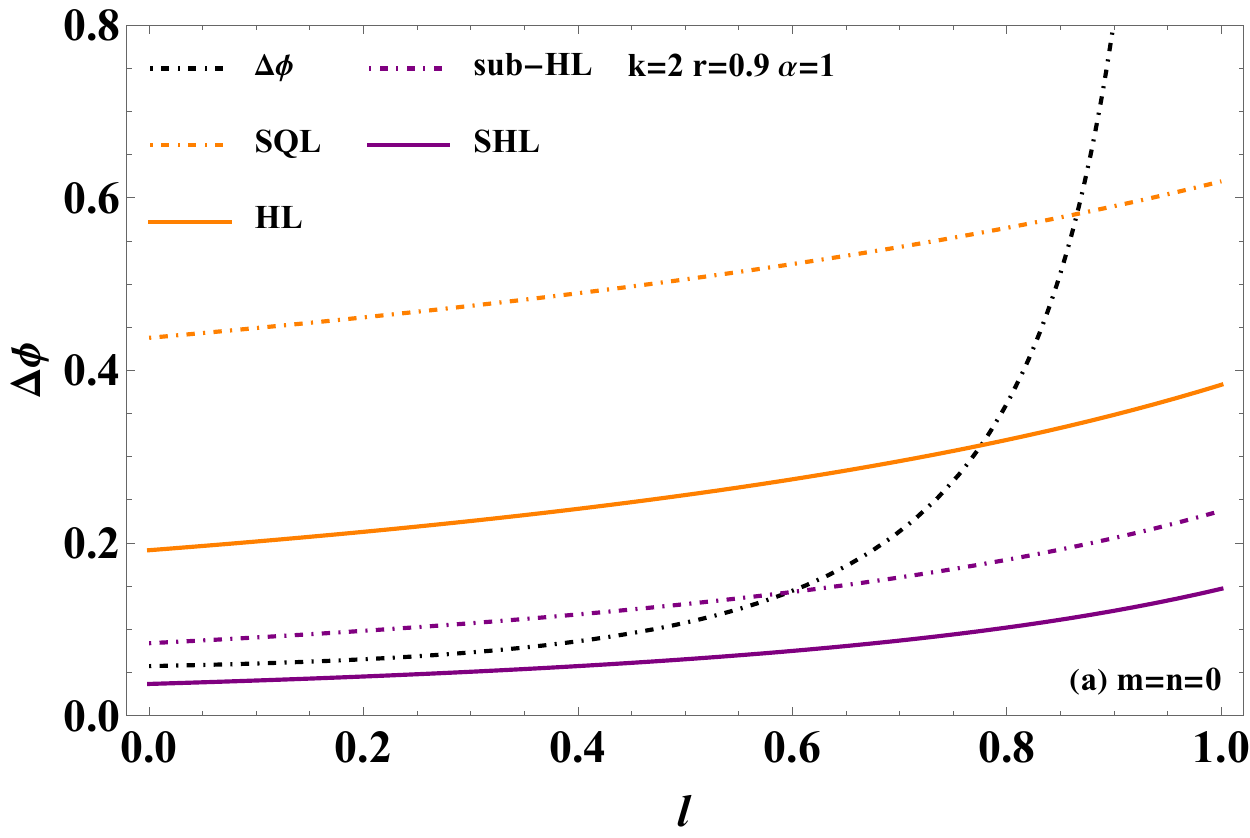} \newline
\includegraphics[width=0.8\columnwidth]{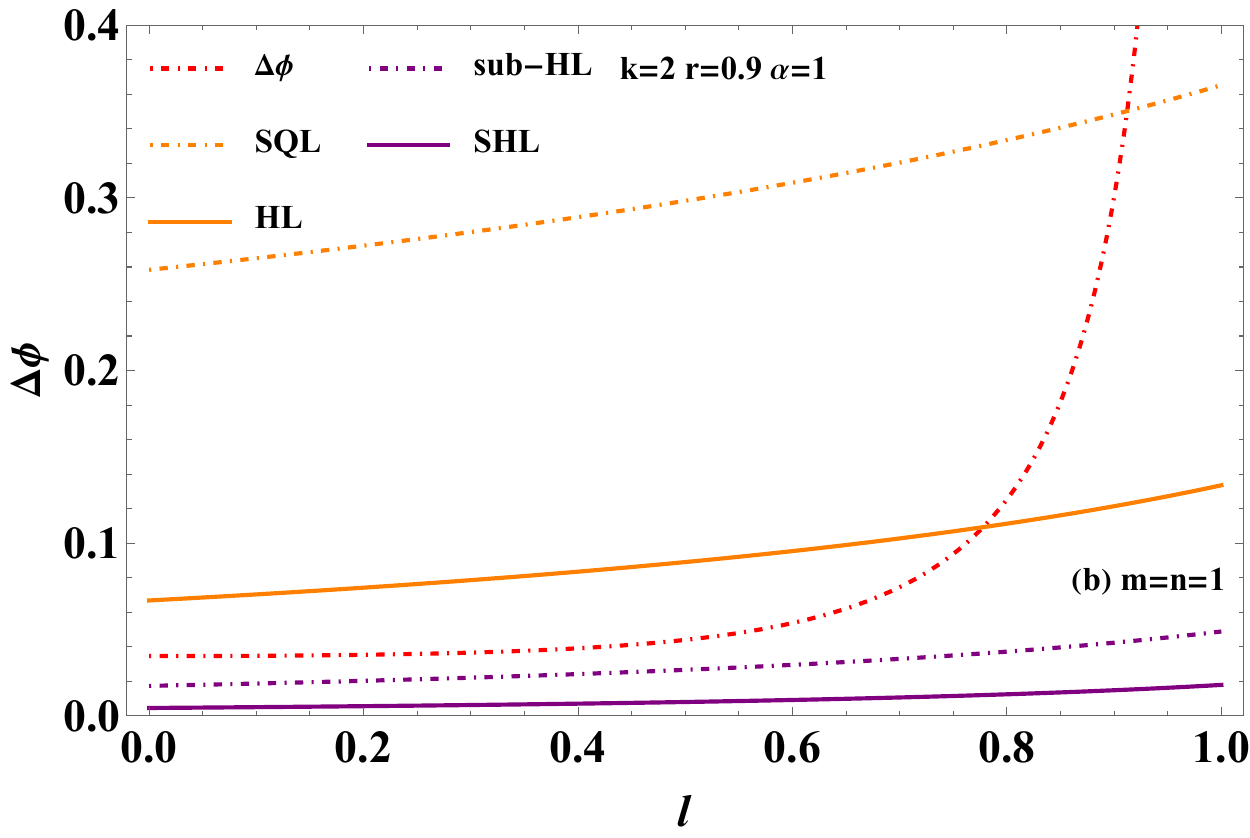} \newline
\includegraphics[width=0.8\columnwidth]{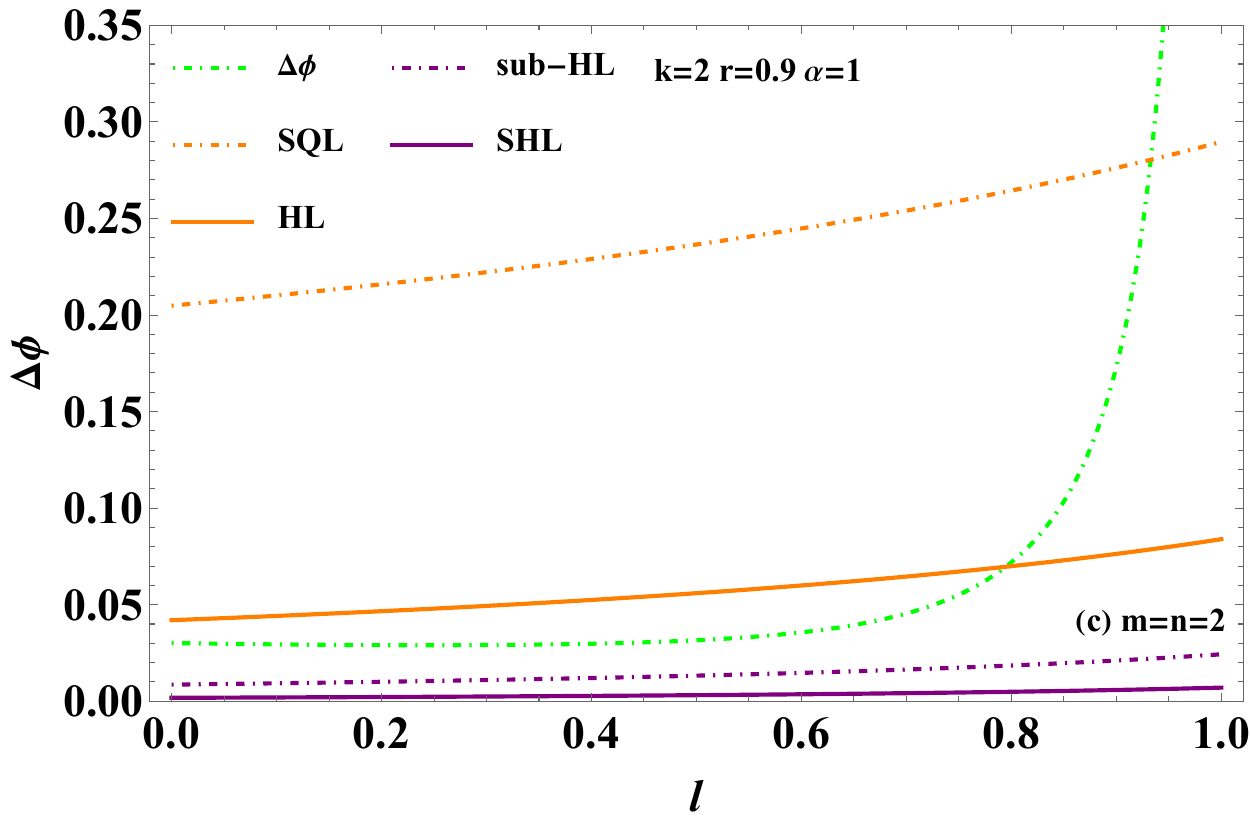}
\caption{Kerr nonlinear phase shift case ($k=2$): phase sensitivity $\Delta%
\protect\phi$ versus loss rate $l$ for (a) $m=n=0$, (b) $m=n=1$, (c) $m=n=2$%
, with fixed phase shift $\protect\phi=3.12$, squeezing parameter $r=0.9$,
and coherent amplitude $\protect\alpha=1$. SQL, HL, sub-HL, and SHL are
shown for comparison.}
\label{Fig5}
\end{figure}

Figs.~\ref{Fig4}(a)--(c) depict linear phase shift sensitivity $\Delta\phi$
as a function of loss rate $l$, compared with SQL and HL. Despite photon
loss, phase sensitivity can still surpass SQL, with photon addition
improving system robustness. Figs.~\ref{Fig5}(a)--(c) further illustrate
Kerr nonlinear phase shift sensitivity versus loss rate $l$. Comparison with
Fig.~\ref{Fig4} clearly shows that Kerr nonlinearity enhances phase
sensitivity by an order of magnitude, enabling exceedance of the HL
threshold even with loss, whereas linear phase shift only surpasses SQL
within limited $l$ ranges. However, with fixed $r=0.9$ and $\alpha=1$, phase
sensitivity for $m=n=0$ can exceed sub-HL, but not for $m=n=1,2$. This may
arise from Kerr nonlinearity sensitivity to other parameters (e.g., phase
shift). This situation indicates that without photon addition, phase
sensitivity can exceed sub-HL within certain ranges, which is relatively
favorable for Kerr nonlinearity.

\section{Quantum Fisher Information}

\subsection{QFI of KMZI in Ideal Case}

Quantum Fisher information (QFI) quantifies the maximum information about
phase shift $\phi$ that a quantum state can convey after passing through a
phase shifter. In our scheme, employing linear and Kerr nonlinear phase
shifters within the MZI leads to two corresponding QFI representations,
denoted $F_{k}$ ($k=1,2$). QFI is independent of any specific detection
scheme and provides an upper bound on classical Fisher information
obtainable through $\phi$ estimation. The optimal error bound for phase
sensitivity, known as the quantum Cram\'er--Rao bound (QCRB), is given by
\cite{44,45,46}:
\begin{equation}
\Delta \phi_{\text{QCRB}_{k}} = \frac{1}{\sqrt{F_{k}}}.  \label{eq:QCRB}
\end{equation}
For pure states, QFI under ideal conditions ($l=0$) can be calculated as
\cite{47,48}:
\begin{equation}
F_{k} = 4 \bigl[ \langle \psi_{\phi}^{\prime }| \psi_{\phi}^{\prime }\rangle
- \lvert \langle \psi_{\phi}^{\prime }| \psi_{\phi} \rangle \rvert^{2} \bigr]%
,  \label{eq:QFI_pure}
\end{equation}
where $\lvert \psi_{\phi} \rangle = U(\phi,k) B_{1} \lvert \psi \rangle_{%
\text{in}}$ represents the quantum state before BS2 in the lossless KMZI,
and $\lvert \psi_{\phi}^{\prime }\rangle = \partial \lvert \psi_{\phi}
\rangle / \partial \phi$. For both linear ($k=1$) and Kerr nonlinear ($k=2$)
phase shifts, QFI can be further simplified as \cite{40}:
\begin{equation}
F_{k} = 4 \langle \Delta^{2} n_{a}^{k} \rangle = 4 \bigl[ \langle
(n_{a}^{k})^{2} \rangle - \langle n_{a}^{k} \rangle^{2} \bigr],
\label{eq:QFI_simple}
\end{equation}
where $\langle \cdot \rangle = \langle \Psi \rvert \cdot \lvert \Psi \rangle$%
, with $\lvert \Psi \rangle = B_{1} \lvert \text{in} \rangle$. Here, $%
n_{a}^{k} = (a^{\dagger}a)^{k}$ ($k=1,2$). To facilitate QFI computation,
the normal ordering form of $n_{a}^{w}$ is obtained via the operator
identity:
\begin{eqnarray}
n_{a}^{w} &=& \frac{\partial^{w}}{\partial x^{w}} \exp\bigl[x a^{\dagger}a%
\bigr] \Big|_{x=0}  \notag \\
&=& \frac{\partial^{w}}{\partial x^{w}} \colon\! \exp\bigl[(e^{x}-1)
a^{\dagger}a\bigr] \!\colon \Big|_{x=0}.  \label{eq:norm_order}
\end{eqnarray}
Since $a^{\dagger}a$ commutes with the phase shifter operator and losses are
neglected here, utilizing the normal ordering of $n_{a}^{w}$ allows deriving
the QFI for $k=1$ using Eq.~(B5) with $l=0$:
\begin{equation}
F_{1} = 4 \bigl[ D_{1}(2,0,2,0) + D_{1}(1,0,1,0) - D_{1}(1,0,1,0)^{2} \bigr],
\label{eq:F1}
\end{equation}
while QFI for $k=2$ becomes:
\begin{eqnarray}
F_{2} &=& 4 \bigl[ D_{1}(4,0,4,0) + 6 D_{1}(3,0,3,0)  \notag \\
&& + 7 D_{1}(2,0,2,0) + D_{1}(1,0,1,0)  \notag \\
&& - \bigl(D_{1}(2,0,2,0) + D_{1}(1,0,1,0)\bigr)^{2} \bigr].  \label{eq:F2}
\end{eqnarray}

\begin{figure*}[t]
\centering
\includegraphics[width=0.49\textwidth]{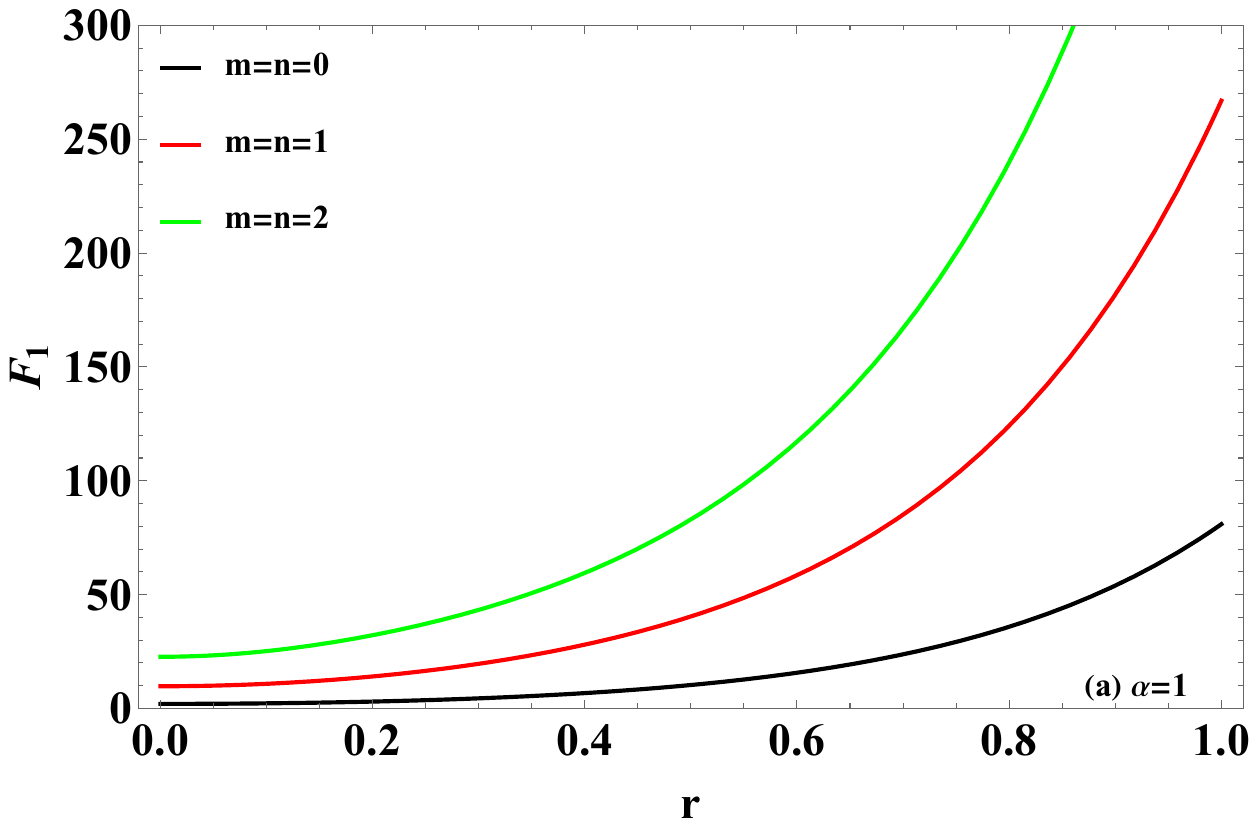} %
\includegraphics[width=0.49\textwidth]{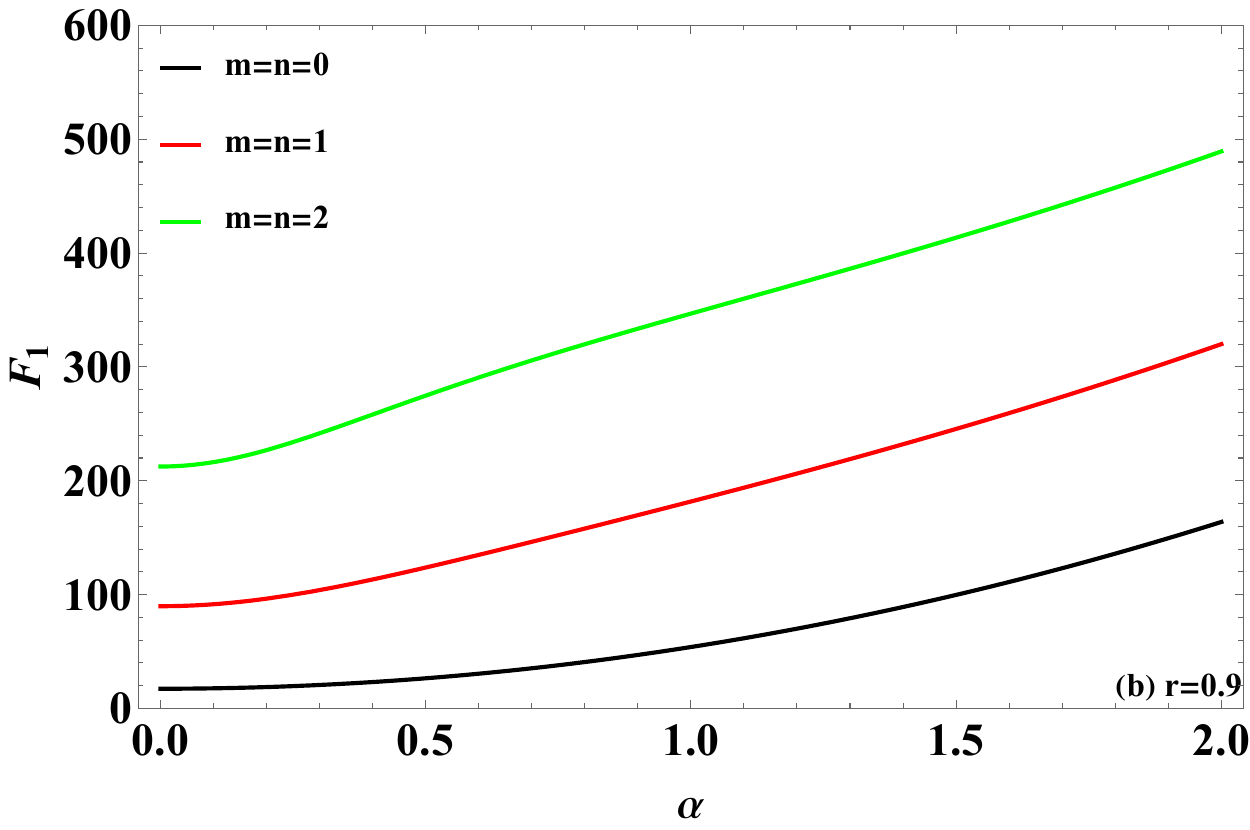} \newline
\includegraphics[width=0.49\textwidth]{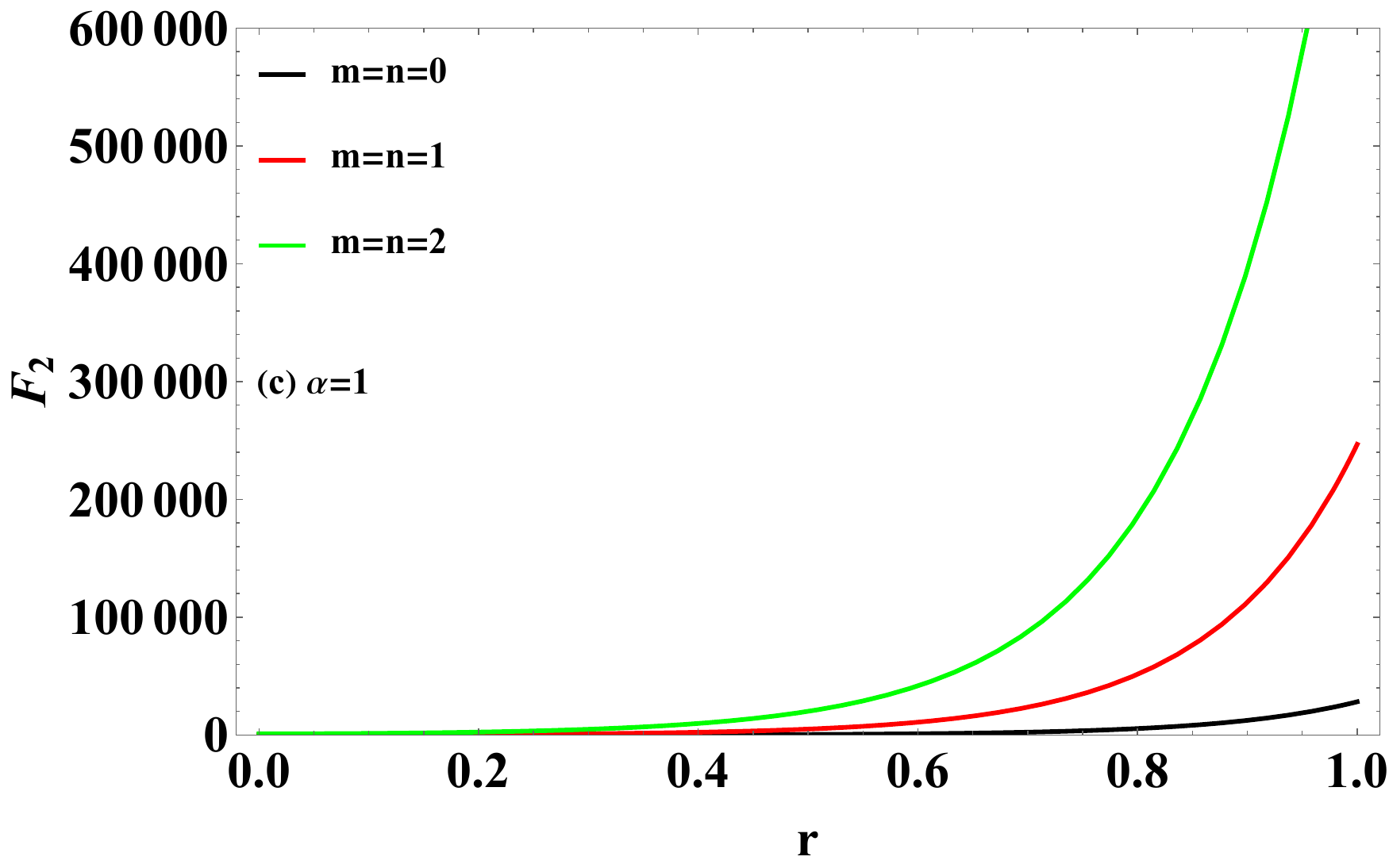} %
\includegraphics[width=0.49\textwidth]{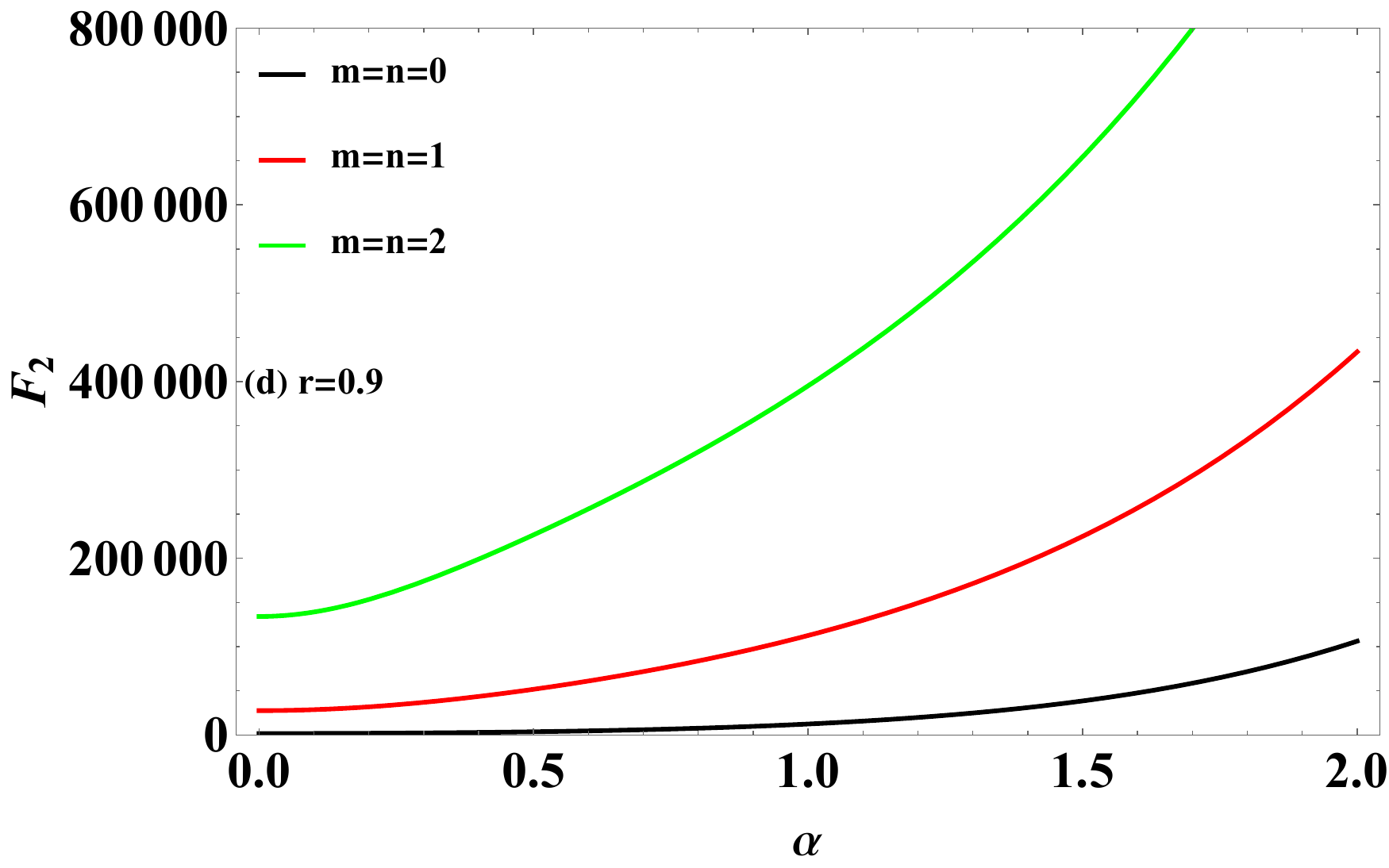}
\caption{For $k=1$ case: (a) QFI $F_{1}$ versus squeezing parameter $r$ with
coherent amplitude $\protect\alpha=1$; (b) $F_{1}$ versus $\protect\alpha$
with $r=0.9$. For $k=2$ case: (c) QFI $F_{2}$ versus $r$ with $\protect\alpha%
=1$; (d) $F_{2}$ versus $\protect\alpha$ with $r=0.9$.}
\label{Fig6}
\end{figure*}

As evident from Fig.~\ref{Fig6}, increasing photon-addition order
significantly enhances QFI. Furthermore, comparing Figs.~\ref{Fig6}(a)--(b)
with Figs.~\ref{Fig6}(c)--(d) clearly shows that the $k=2$ case yields
larger QFI values. This demonstrates that both photon-addition operations
and Kerr nonlinear phase shifts effectively improve QFI.

\subsection{Impacts of Photon Losses on QFI}

\begin{figure}[t]
\centering
\includegraphics[width=0.7\columnwidth]{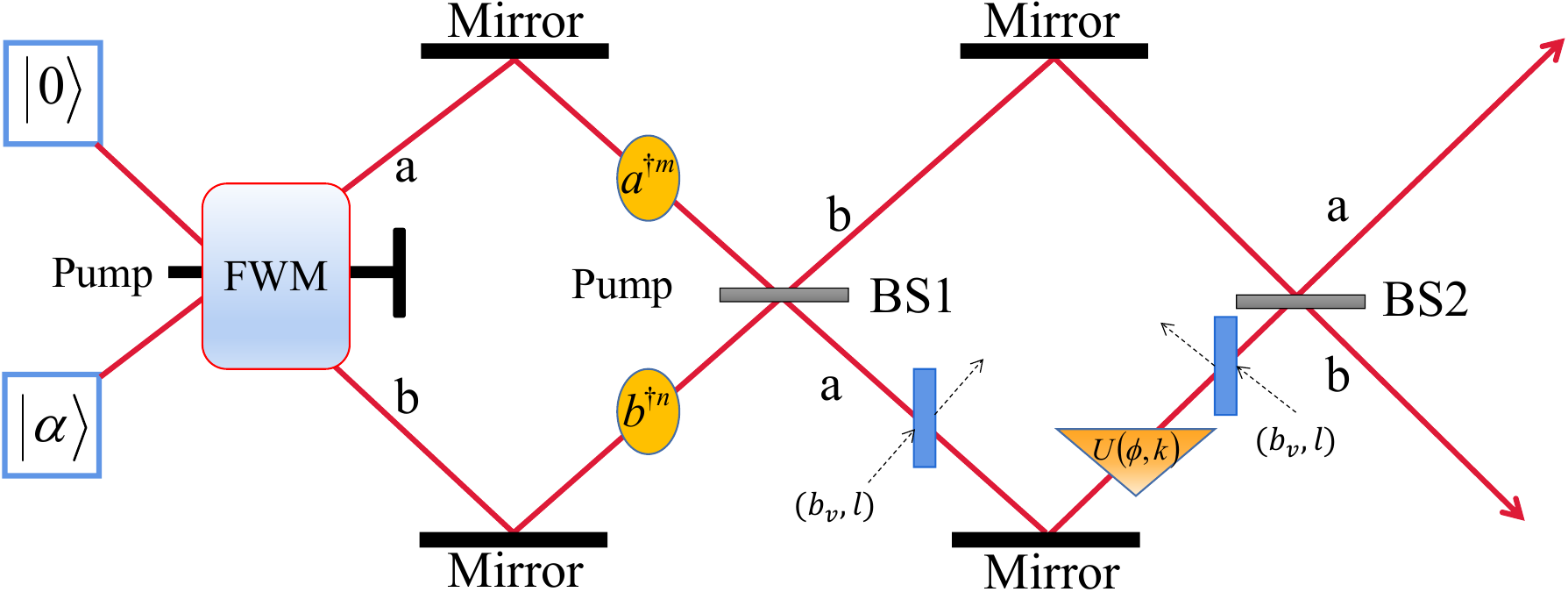}
\caption{Theoretical model of lossy KMZI using PA-TMSCS as input. Two
virtual beam splitters, positioned before and after the phase shifter,
simulate photon losses. Parameter $l$ represents beam splitter reflectivity
(loss rate), while $b_{v}$ denotes the vacuum operator.}
\label{Fig7}
\end{figure}

\subsubsection{Effects of Photon Loss on QFI in Linear Phase Shift Case ($%
k=1 $)}

Practical measurements involve photon losses due to vacuum noise, affecting
QFI. For the initial pure state $\lvert \Psi_{S} \rangle$ ($\lvert \Psi_{S}
\rangle = B_{1} \lvert \psi \rangle_{\text{in}}$) in a probe system $S$ with
a lossy MZI, we introduce environment $E$ orthogonal states $\lvert j_{E}
\rangle$ and Kraus operators $\hat{\Pi}_{j}(\phi)$ to characterize $\lvert
\Psi_{S} \rangle$ evolution through the phase shifter with photon losses.
Fig.~\ref{Fig7} illustrates the photon loss within the MZI, providing a
visual representation of the model structure under investigation. Following
Escher \emph{et al.}'s methods for calculating QFI in open quantum systems
\cite{49}, the quantum states $\lvert \Psi_{S} \rangle$ and environment
vacuum noise $\lvert 0_{E} \rangle$, after unitary evolution $U_{S+E}(\phi)$%
, can be expressed in the extended $S+E$ space:
\begin{eqnarray}
\lvert \Psi_{S+E} \rangle &=& U_{S+E}(\phi) \lvert \Psi_{S} \rangle \lvert
0_{E} \rangle  \notag \\
&=& \sum_{j=0}^{\infty} \hat{\Pi}_{j}(\phi) \lvert \Psi_{S} \rangle \lvert
j_{E} \rangle.  \label{eq:purified_state}
\end{eqnarray}
Although photon losses transform $\lvert \Psi_{S} \rangle$ into a mixed
state, this approach allows treating $\lvert \Psi_{S+E} \rangle$ in $S+E$ as
a pure state. For the entire purified system, QFI under photon losses can be
expressed as:
\begin{equation}
F_{L_{1}} \leqslant C_{Q}\bigl[ \lvert \Psi_{S} \rangle, \hat{\Pi}_{j}(\phi) %
\bigr] = 4 \bigl[ \langle \hat{H}_{1} \rangle_{S} - \lvert \langle \hat{H}%
_{2} \rangle_{S} \rvert^{2} \bigr],  \label{eq:CQ_bound}
\end{equation}
where the lower bound of $C_{Q}[\lvert \Psi_{S} \rangle, \hat{\Pi}%
_{j}(\phi)] $ is shown to be the QFI for the reduced system \cite{49}, and $%
\hat{H}_{1,2} $ are Hermitian operators defined as:
\begin{eqnarray}
\hat{H}_{1} &=& \sum_{j=0}^{\infty} \frac{d\hat{\Pi}_{j}^{\dagger}(\phi)}{%
d\phi} \frac{d\hat{\Pi}_{j}(\phi)}{d\phi},  \label{eq:H1_def} \\
\hat{H}_{2} &=& i \sum_{j=0}^{\infty} \frac{d\hat{\Pi}_{j}^{\dagger}(\phi)}{%
d\phi} \hat{\Pi}_{j}(\phi),  \label{eq:H2_def}
\end{eqnarray}
with $\langle \cdot \rangle_{S}$ denoting expectation value $\langle
\Psi_{S} \rvert \cdot \lvert \Psi_{S} \rangle$, representing the average of $%
\hat{H}_{1,2}$. Thus, $F_{L_{1}} = C_{Q\min}$, and for linear phase shift ($%
k=1$), the Kraus operator is:
\begin{equation}
\hat{\Pi}_{j}(\phi) = \sqrt{\frac{l^{j}}{j!}} e^{i\phi (a^{\dagger}a -
\gamma j)} (1-l)^{\frac{a^{\dagger}a}{2}} a^{j},  \label{eq:kraus_linear}
\end{equation}
where $l$ is the loss rate, and $\gamma = 0$ or $-1$ corresponds to photon
losses before or after the linear phase shifter, respectively (Fig.~\ref%
{Fig7}). Optimizing $\gamma$ yields $C_{Q\min}$. Thus, QFI with photon
losses becomes \cite{49}:
\begin{equation}
F_{L_{1}} = \frac{4 F_{1} (1-l) \langle n_{a} \rangle}{l F_{1} + 4 (1-l)
\langle n_{a} \rangle}.  \label{eq:F1_loss}
\end{equation}

\subsubsection{Effects of Photon Loss on QFI in Kerr Nonlinear Phase Shift
Case ($k=2$)}

Extending the previous approach, we investigate QFI under photon losses for
Kerr nonlinear phase shift ($k=2$). The general Kraus operator form,
incorporating Kerr nonlinear phase shift, is defined as \cite{40}:
\begin{equation}
\hat{\Pi}_{j}(\phi) = \sqrt{\frac{l^{j}}{j!}} e^{i\phi [ (a^{\dagger}a)^{2}
- 2\mu_{1} a^{\dagger}a j - \mu_{2} j^{2} ]} (1-l)^{\frac{a^{\dagger}a}{2}}
a^{j},  \label{eq:kraus_kerr}
\end{equation}
where parameters $\mu_{1} = \mu_{2} = 0$ or $-1$ correspond to photon losses
before or after the Kerr nonlinear phase shifter (Fig.~\ref{Fig7}).
Referring to Eq.~(\ref{eq:CQ_bound}), we derive (see Appendix~D for
derivation):
\begin{eqnarray}
F_{L_{2}} &\leqslant& C_{Q}\bigl[ \lvert \Psi_{S} \rangle, \hat{\Pi}%
_{j}(\phi) \bigr]  \notag \\
&=& 4 \bigl[ K_{1}^{2} \langle \Delta^{2} n_{a}^{2} \rangle_{S} - K_{2}
\langle n_{a}^{3} \rangle_{S}  \notag \\
&& + K_{3} \langle n_{a}^{2} \rangle_{S} - K_{4} \langle n_{a} \rangle_{S}
\notag \\
&& - K_{5} \langle n_{a}^{2} \rangle_{S} \langle n_{a} \rangle_{S} - K_{6}
\langle n_{a} \rangle_{S}^{2} \bigr],  \label{eq:CQ_kerr}
\end{eqnarray}
where $K_{i}$ ($i=1,2,3,4,5,6$) are detailed in Appendix~D. To determine QFI
$F_{QL}$ under photon losses for Kerr nonlinear phase shift ($k=2$),
substitute optimal values $\mu_{1\text{opt}}$ and $\mu_{2\text{opt}}$ into $%
C_{Q}$ to find $C_{Q\min}$. Specific expressions for $\mu_{1\text{opt}}$ and
$\mu_{2\text{opt}}$ are provided in Appendix~D. Similar to Eq.~(\ref{eq:QCRB}%
), $\Delta\phi_{\text{QBL}_{k}}$ under photon losses for $k=1$ and $k=2$ is
derived from Eqs.~(\ref{eq:F1_loss}) and (\ref{eq:CQ_kerr}), respectively,
as $\Delta\phi_{\text{QBL}_{k}} = 1/\sqrt{F_{L_{k}}}$.

\begin{figure}[t]
\centering
\includegraphics[width=0.8\columnwidth]{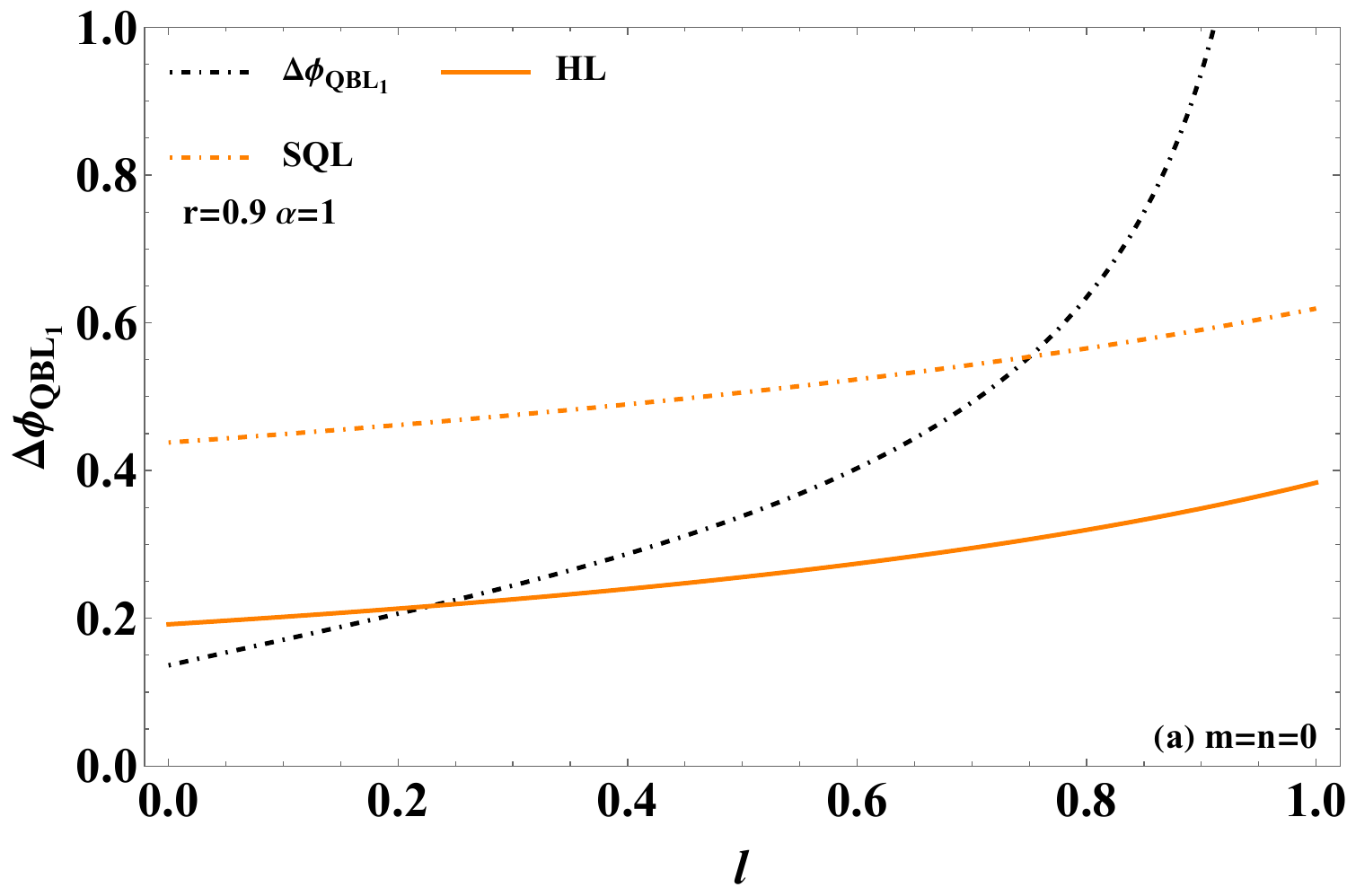} \newline
\includegraphics[width=0.8\columnwidth]{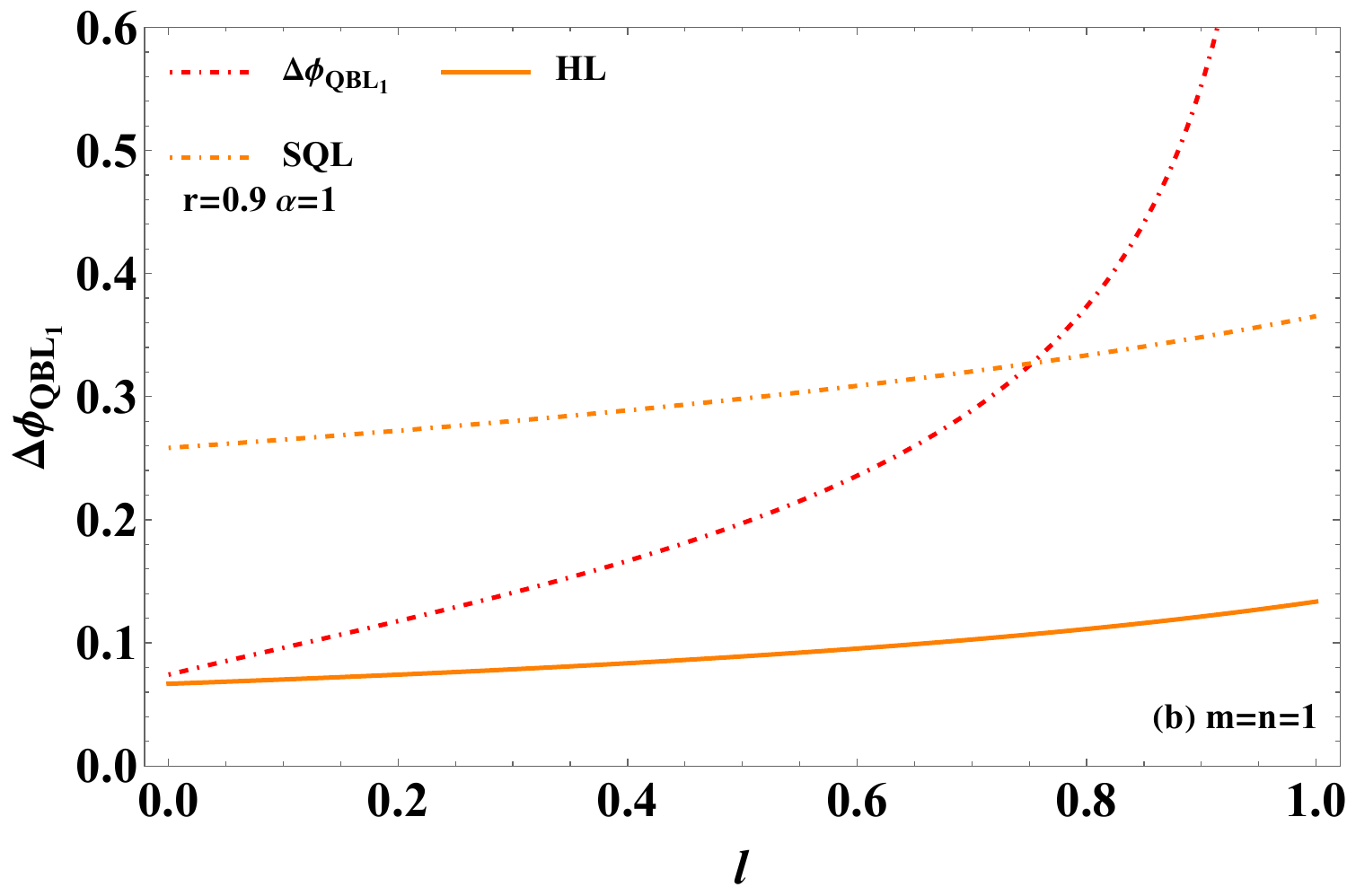} \newline
\includegraphics[width=0.8\columnwidth]{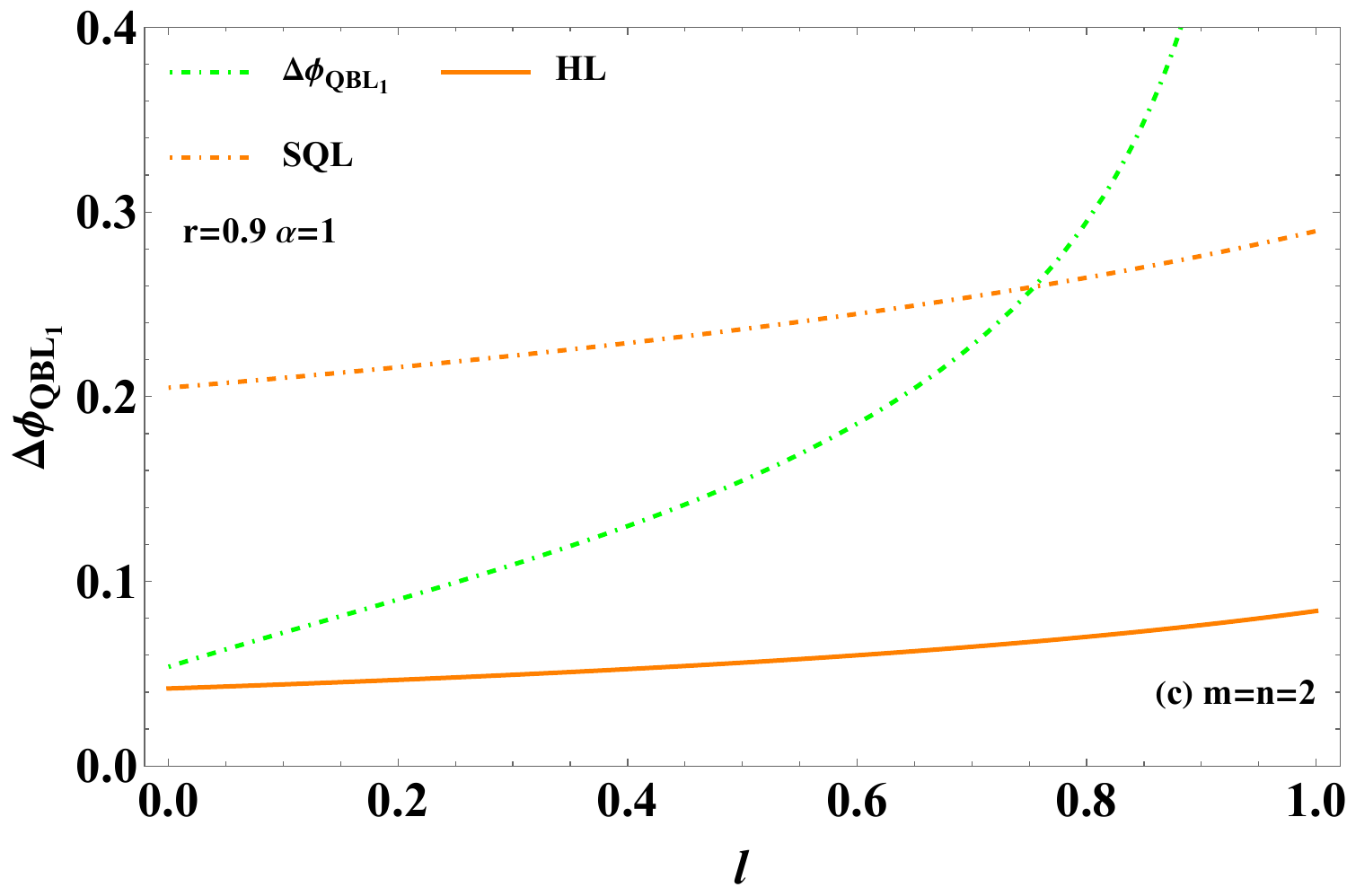}
\caption{For $k=1$ case: QCRB $\Delta\protect\phi_{\text{QBL}_{1}}$ versus
loss rate $l$ for (a) $m=n=0$, (b) $m=n=1$, (c) $m=n=2$, with fixed
squeezing parameter $r=0.9$ and coherent amplitude $\protect\alpha=1$. SQL
and HL are shown for comparison.}
\label{Fig8}
\end{figure}

As shown in Fig.~\ref{Fig8}, despite photon loss, QCRB can still be improved
through photon addition and can surpass SQL across wide loss rate $l$
ranges, even reaching HL for small $l$.

\begin{figure}[t]
\centering
\includegraphics[width=0.8\columnwidth]{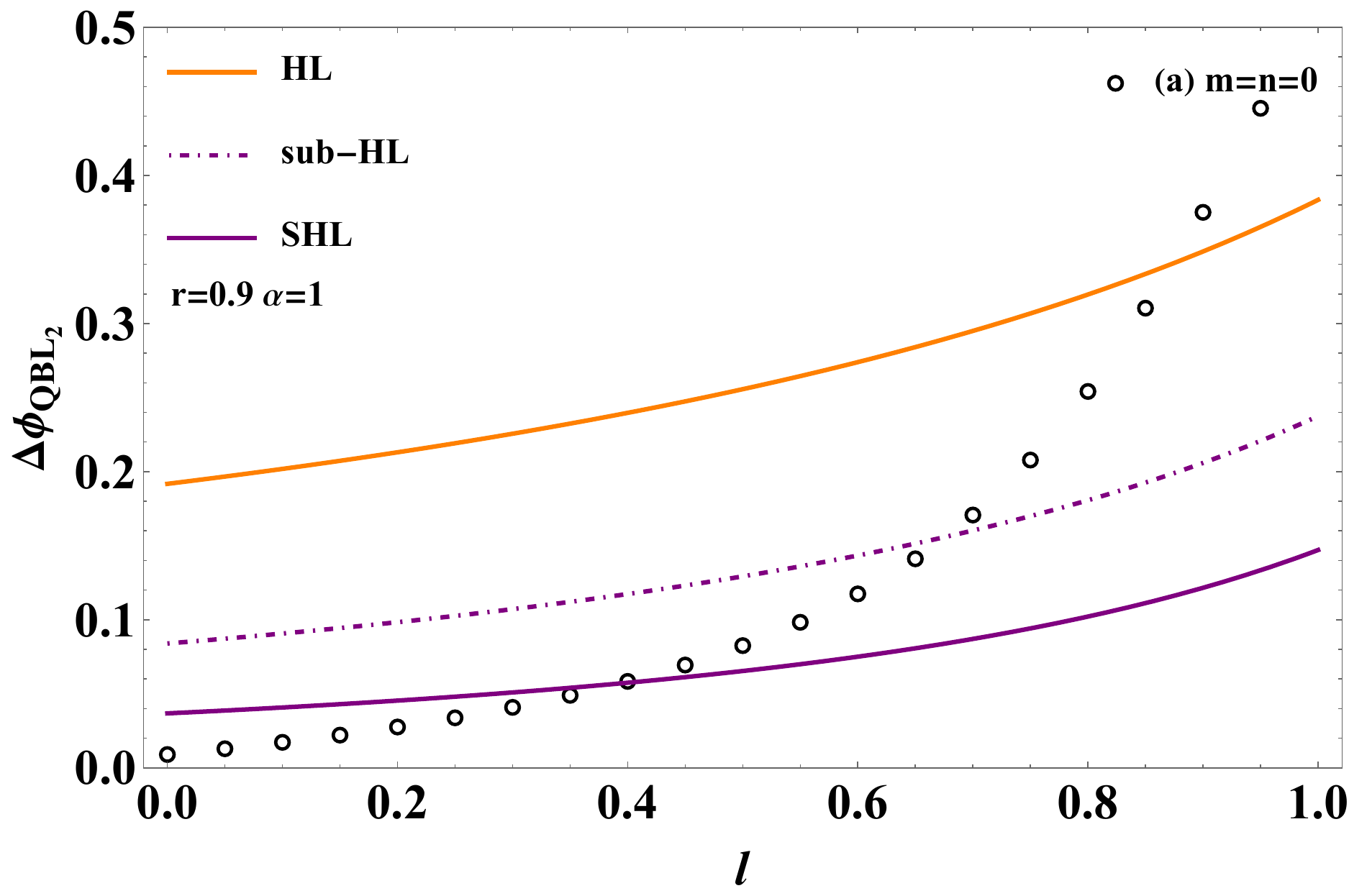} \newline
\includegraphics[width=0.8\columnwidth]{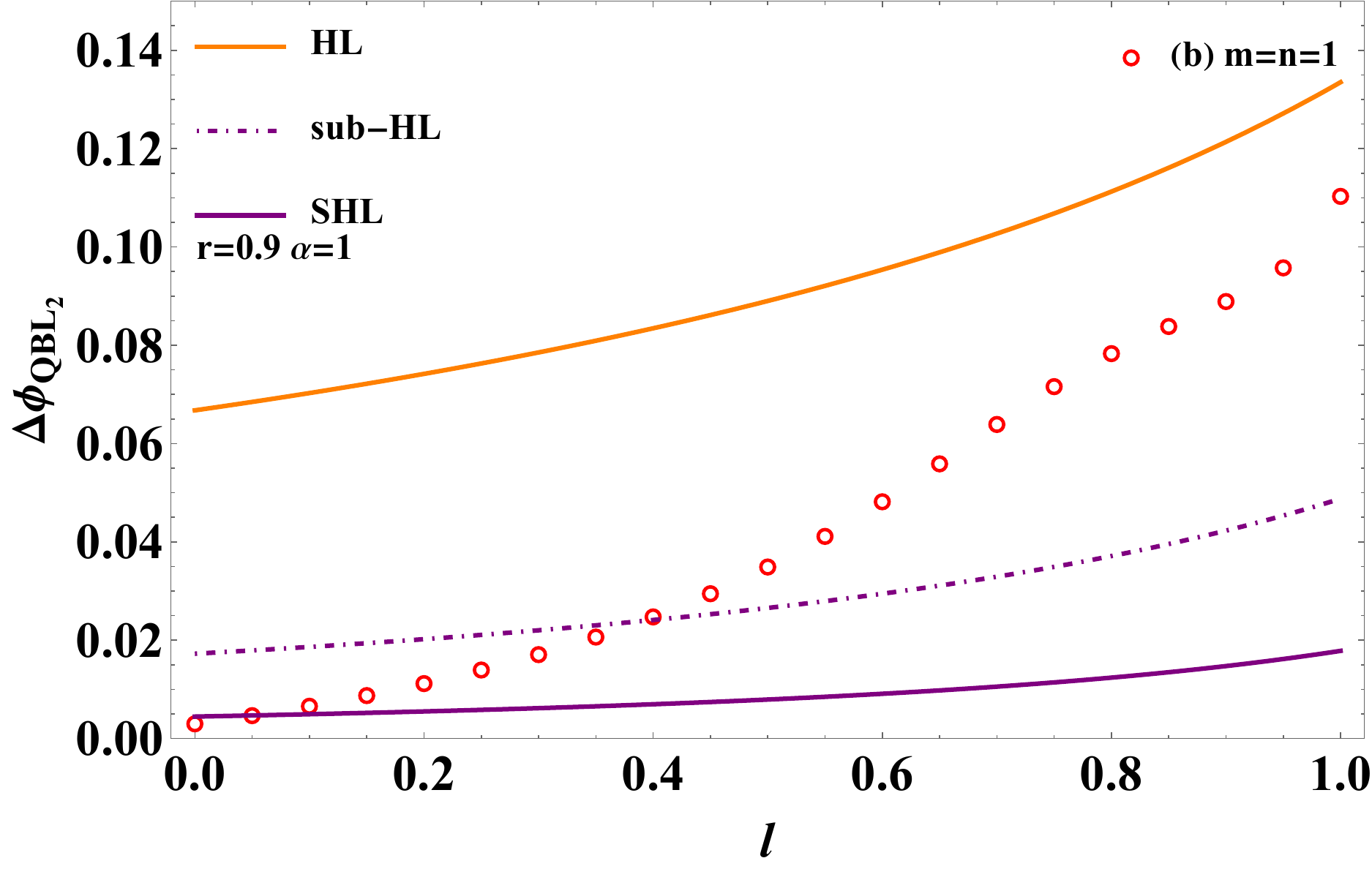} \newline
\includegraphics[width=0.8\columnwidth]{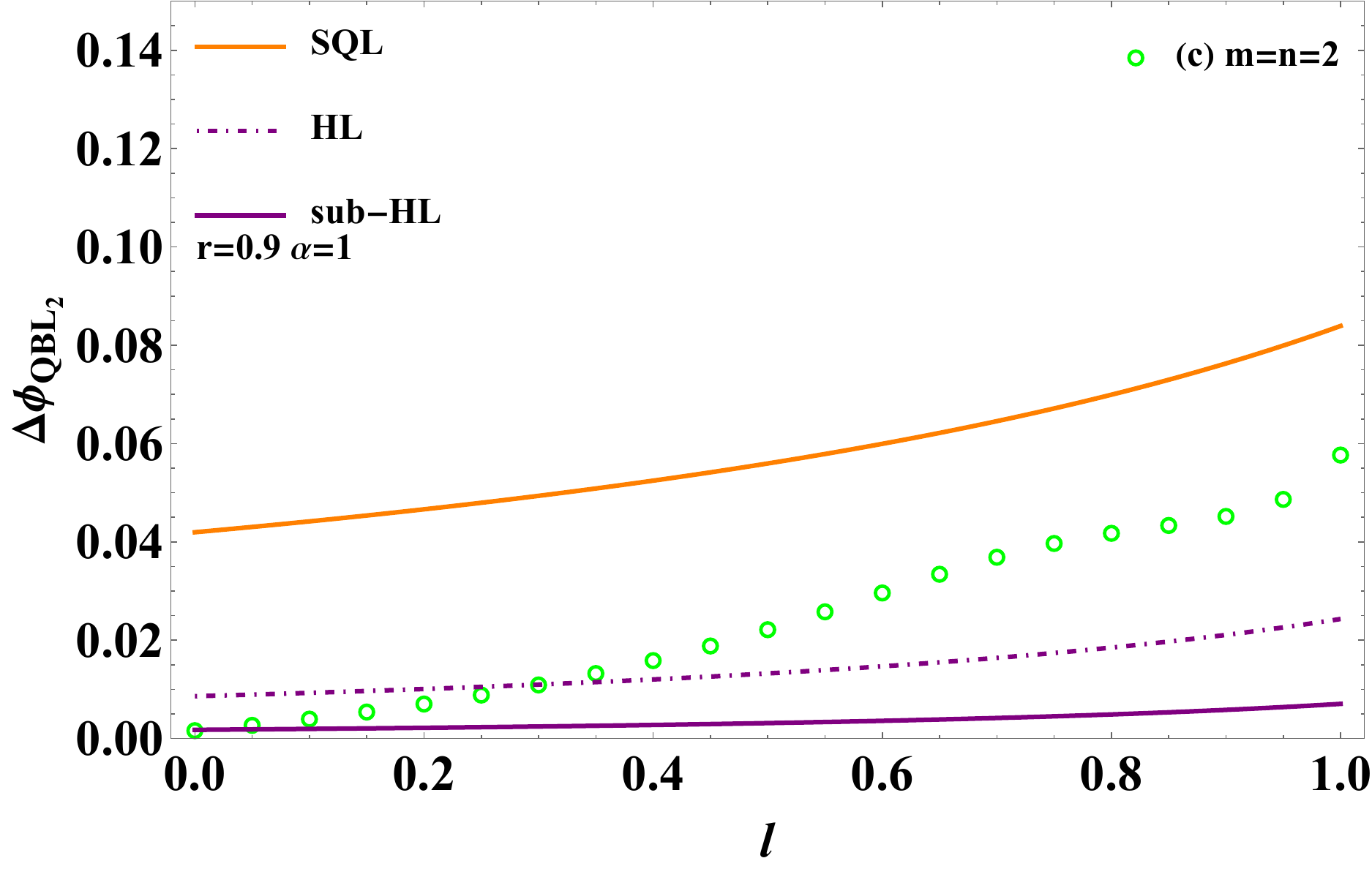}
\caption{For $k=2$ case: QCRB $\Delta\protect\phi_{\text{QBL}_{2}}$ versus
loss rate $l$ for (a) $m=n=0$, (b) $m=n=1$, (c) $m=n=2$, with fixed
squeezing parameter $r=0.9$ and coherent amplitude $\protect\alpha=1$. SQL,
HL, sub-HL, and SHL are shown for comparison.}
\label{Fig9}
\end{figure}

Comparing Fig.~\ref{Fig8} with Fig.~\ref{Fig9} reveals that Kerr nonlinear
phase shift ($k=2$) significantly enhances QCRB. Despite losses, QCRB
improvement through photon addition persists, with QCRB breaking through HL
and sub-HL across wide $l$ ranges, even exceeding SHL for small $l$. This
demonstrates that although photon loss degrades QCRB, both photon addition
and Kerr nonlinear phase shift can mitigate this degradation, thereby
enhancing system robustness.

\section{Conclusion}

We have proposed and analyzed an enhanced phase estimation scheme combining
PA-TMSCS with a KMZI. Our results demonstrate that this approach
systematically improves phase sensitivity and QFI, enabling the surpassing
of fundamental precision limits. Specifically, the scheme exceeds the SQL
and approaches the HL for linear phase shifts, while the Kerr nonlinearity
facilitates scaling beyond the SHL. Moreover, the integrated design exhibits
enhanced robustness against photon loss, a critical attribute for practical
implementations.

This work elucidates the synergistic mechanism between non-Gaussian photon
addition and Kerr nonlinearity, which concurrently boosts measurement
precision and resilience. The findings establish a valuable theoretical
framework for developing practical quantum-enhanced metrology protocols,
with potential applications in ultra-precise sensing. Future work may focus
on identifying optimal operating parameters and experimental realizations in
physical platforms suitable for nonlinear quantum optics.

\section*{Acknowledgments}

This work is supported by the National Natural Science Foundation of China
(Grants No. 12564049 and No. 12104195) and the Jiangxi Provincial Natural
Science Foundation (Grants No. 20242BAB26009 and 20232BAB211033), Jiangxi
Provincial Key Laboratory of Advanced Electronic Materials and Devices
(Grant No. 2024SSY03011), Jiangxi Civil-Military Integration Research
Institute (Grant No. 2024JXRH0Y07), as well as the Science and Technology
Project of Jiangxi Provincial Department of Education (Grant No. GJJ2404102).

\appendix

\section{Derivation of Transformation Relation for Kerr Nonlinear Phase
Shifter Operator}

\label{app:A}

In the scheme proposed in this paper, the Kerr nonlinear phase shifter
significantly enhances phase measurement accuracy. The associated
calculations can be efficiently performed using the transformation relation
of the equivalent operator of the Kerr nonlinear phase shifter, $U(\phi,2) =
e^{i\phi (a^{\dagger}a)^{2}}$, as shown in our study. Therefore, this
appendix provides a detailed derivation of the transformation relation $%
U^{\dagger}(\phi,2) a U(\phi,2) = e^{i\phi} e^{i\phi (2 a^{\dagger}a)} a$ in
Eq.~(\ref{eq:kerr_trans}). Here, we assume the operator corresponding to the
transformation relation of $U(\phi,2)$ is $O$, i.e., $O =
U^{\dagger}(\phi,2) a U(\phi,2)$. The matrix of $O$ in the Fock state space
can be represented by $O_{n,n^{\prime }} = \langle n \rvert O \lvert
n^{\prime }\rangle$, and its corresponding expansion is given by
\begin{equation}
O = \sum_{n,n^{\prime }=0}^{\infty} \lvert n \rangle \langle n^{\prime
}\rvert O_{n,n^{\prime }}.  \tag{A1}
\end{equation}
By utilizing $\langle n \rvert a^{\dagger}a = \langle n \rvert n$ and $%
\langle n \rvert a = \langle n+1 \rvert \sqrt{n+1}$, the matrix element $%
O_{n,n^{\prime }}$ in the aforementioned equation can be further calculated
as
\begin{equation}
O_{n,n^{\prime }} = \sqrt{n+1} e^{i\phi (2n+1)} \delta_{n+1,n^{\prime }},
\tag{A2}
\end{equation}
and by substituting it into Eq.~(A1), the transformation relation of $%
U(\phi,2)$ can be deduced as follows
\begin{align}
O &= e^{i\phi} \sum_{n=0}^{\infty} \lvert n \rangle \langle n+1 \rvert \sqrt{%
n+1} e^{i2n\phi}  \notag \\
&= e^{i\phi} \sum_{n=0}^{\infty} \lvert n \rangle \langle n \rvert
e^{i2n\phi} a = e^{i\phi} e^{i\phi (2 a^{\dagger}a)} a,  \tag{A3}
\end{align}
where we have used the completeness of the Fock state $\sum_{n=0}^{\infty}
\lvert n \rangle \langle n \rvert = 1$.

\section{Specific Derivations and Expressions for $\langle I_{D} \rangle$
and $\langle I_{D}^{2} \rangle$ under $k=1$ Condition}

\label{app:B}

In order to obtain the mean values $\langle I_{D} \rangle$ and $\langle
I_{D}^{2} \rangle$ in Eq.~(\ref{eq:phase_sensitivity}) in a simple way, the
general formula is introduced, i.e.,
\begin{align}
& {}_{f}\langle 0 \rvert \langle \text{in} \rvert B_{1}^{\dagger}
B_{f}^{\dagger} e^{-i\phi a^{\dagger}a}  \notag \\
& \times \bigl( a^{\dagger m_{1}} b^{\dagger n_{1}} a^{m_{2}} b^{n_{2}} %
\bigr) e^{i\phi a^{\dagger}a} B_{f} B_{1} \lvert \text{in} \rangle \lvert 0
\rangle_{f}  \notag \\
& = D_{1}(m_{1},n_{1},m_{2},n_{2}).  \tag{B1}
\end{align}
The quantum state used in the above formula to calculate the average value
is the state after passing through the phase shifter, and the output state
still needs to pass through BS2. Thus using the transformation relation
Eqs.~(\ref{eq:BS2_trans}) and (B1) one can obtain $\langle I_{D} \rangle$
and $\langle I_{D}^{2} \rangle$ respectively as
\begin{equation}
\langle I_{D} \rangle = i \bigl[ D_{1}(1,0,0,1) - D_{1}(0,1,1,0) \bigr],
\tag{B2}
\end{equation}
and
\begin{align}
\langle I_{D}^{2} \rangle &= D_{1}(0,1,0,1) + D_{1}(1,0,1,0) + 2
D_{1}(1,1,1,1)  \notag \\
&\quad - D_{1}(2,0,0,2) - D_{1}(0,2,2,0).  \tag{B3}
\end{align}
In order to facilitate the derivation of general formula Eq.~(B1), we use
the following partial derivative form for representation,
\begin{align}
& D_{1}(m_{1},n_{1},m_{2},n_{2})  \notag \\
& = {}_{f}\langle 0 \rvert \langle \text{in} \rvert B_{1}^{\dagger}
B_{f}^{\dagger} e^{-i\phi a^{\dagger}a} \frac{%
\partial^{m_{1}+n_{1}+m_{2}+n_{2}}}{\partial x_{1}^{m_{1}} \partial
y_{1}^{n_{1}} \partial x_{2}^{m_{2}} \partial y_{2}^{n_{2}}}  \notag \\
& \quad \exp\bigl[ a^{\dagger} x_{1} + b^{\dagger} y_{1} + a x_{2} + b y_{2} %
\bigr] \Big|_{x_{1}=y_{1}=x_{2}=y_{2}=0}  \notag \\
& \quad \times e^{i\phi a^{\dagger}a} B_{f} B_{1} \lvert \text{in} \rangle
\lvert 0 \rangle_{f}.  \tag{B4}
\end{align}
According to the transformation relationship of the corresponding operator,
it can be calculated that
\begin{align}
& \Gamma_{1} \Bigl\{ \exp\bigl[ \alpha (t_{1}+t_{2}+A_{1}+A_{4}) \cosh r %
\bigr]  \notag \\
& \times \exp\bigl[ \alpha (\tau_{1}+\tau_{2}+A_{2}+A_{3}) \sinh r \bigr]
\notag \\
& \times \exp\bigl[ (t_{1} \sinh r + A_{2} \cosh r) \tau_{1} \cosh r  \notag
\\
& \quad + (\tau_{2} \sinh r + A_{1} \cosh r) t_{2} \cosh r \bigr]  \notag \\
& \times \exp\bigl[ (t_{1} \cosh r + A_{2} \sinh r)  \notag \\
& \quad \times (A_{3} \sinh r + (t_{2}+A_{4}) \cosh r) \bigr]  \notag \\
& \times \exp\bigl[ (\tau_{2} \cosh r + A_{1} \sinh r)  \notag \\
& \quad \times (A_{4} \sinh r + (\tau_{1}+A_{3}) \cosh r) \bigr] \Bigr\}
\notag \\
& = D_{1}(m_{1},n_{1},m_{2},n_{2}),  \tag{B5}
\end{align}
where
\begin{align}
\Gamma_{1}\{\cdot\} &= \frac{1}{N_{m,n}} \frac{%
\partial^{2m+2n+m_{1}+n_{1}+m_{2}+n_{2}}}{\partial t_{1}^{m} \partial
\tau_{1}^{n} \partial t_{2}^{m} \partial \tau_{2}^{n} \partial x_{1}^{m_{1}}
\partial y_{1}^{n_{1}} \partial x_{2}^{m_{2}} \partial y_{2}^{n_{2}}}
\{\cdot\}  \notag \\
& \quad \Big|_{t_{1}=\tau_{1}=t_{2}=\tau_{2}=x_{1}=y_{1}=x_{2}=y_{2}=0},
\tag{B6}
\end{align}
and
\begin{align}
A_{1} &= \frac{\sqrt{2}}{2} \bigl( \sqrt{1-l} e^{-i\phi} x_{1} + i y_{1} %
\bigr),  \notag \\
A_{2} &= \frac{\sqrt{2}}{2} \bigl( y_{2} - i \sqrt{1-l} e^{i\phi} x_{2} %
\bigr),  \notag \\
A_{3} &= \frac{\sqrt{2}}{2} \bigl( y_{1} + i \sqrt{1-l} e^{-i\phi} x_{1} %
\bigr),  \notag \\
A_{4} &= \frac{\sqrt{2}}{2} \bigl( \sqrt{1-l} e^{i\phi} x_{2} - i y_{2} %
\bigr).  \tag{B7}
\end{align}

\section{Specific Derivations and Expressions for $\langle I_{D} \rangle$
and $\langle I_{D}^{2} \rangle$ under $k=2$ Condition}

\label{app:C}

With respect to operators that are partially obtained from the
transformation relation of BS2, such as $a^{\dagger}b$, it is necessary to
utilize the transformation relation of the Kerr nonlinear phase shift (Eq.~(%
\ref{eq:kerr_trans})) and employ the IWOP technique to facilitate the
calculation of the average values $\langle I_{D} \rangle$ and $\langle
I_{D}^{2} \rangle$. Hence, we construct the general formula
\begin{align}
& D_{2}(x)  \notag \\
& = {}_{f}\langle 0 \rvert \langle \text{in} \rvert B_{1}^{\dagger}
B_{f}^{\dagger} e^{-i\phi (a^{\dagger}a)^{2}} (a^{\dagger}b)^{x} e^{i\phi
(a^{\dagger}a)^{2}} B_{f} B_{1} \lvert \text{in} \rangle \lvert 0 \rangle_{f}
\notag \\
& = \langle \psi \rvert e^{-x^{2} i\phi} a^{\dagger x} e^{-2x i\phi
a^{\dagger}a} b^{x} \lvert \psi \rangle  \notag \\
& = \frac{\partial^{x}}{\partial s^{x}} \langle \psi \rvert e^{-x^{2} i\phi}
\exp[a^{\dagger} b s] e^{-2x i\phi a^{\dagger}a} \lvert \psi \rangle \Big|%
_{s=0},  \tag{C1}
\end{align}
where $x=1,2$ and for the convenience of using IWOP technology, $B_{f} B_{1}
\lvert \text{in} \rangle \lvert 0 \rangle_{f}$ is converted into a quantum
state that represents a coherent state, denoted as $\lvert \psi \rangle$.
Using the following normal ordering form
\begin{equation}
e^{-2x i\phi a^{\dagger}a} = \colon\! \exp\bigl[ (e^{-2x i\phi} - 1)
a^{\dagger}a \bigr] \!\colon,  \tag{C2}
\end{equation}
and integral formula
\begin{equation}
\int \frac{d^{2}z}{\pi} e^{\zeta \lvert z \rvert^{2} + \xi z + \eta z^{\ast}
+ f z^{2} + g z^{\ast 2}} = \frac{e^{\frac{-\zeta \xi \eta + \xi^{2} g +
\eta^{2} f}{\zeta^{2} - 4fg}}}{\sqrt{\zeta^{2} - 4fg}},  \tag{C3}
\end{equation}
we can compute $D_{2}(x)$ as
\begin{equation}
D_{2}(x) = \frac{\exp\bigl[ -x^{2} i\phi - \alpha^{2} + M_{0} + \frac{-M_{1}
M_{2} M_{3} + M_{4} M_{2}^{2} + M_{5} M_{3}^{2}}{M_{1}^{2} - 4 M_{4} M_{5}} %
\bigr]}{\cosh^{2} r \, N_{m,n} \sqrt{M_{1}^{2} - 4 M_{4} M_{5}}},  \tag{C4}
\end{equation}
where
\begin{align}
M_{0} &= V_{5} \bigl( -i \tanh r (V_{1}+V_{2})(V_{1}-V_{2}) V_{5}  \notag \\
& \quad - i (V_{1}-V_{2}) \tau_{1} + (V_{1}+V_{2}) V_{4} \bigr),  \notag \\
M_{1} &= \tanh^{2} r \frac{4V_{1}^{2} + 1 + (1-l) s^{2}}{2} - 1,  \notag \\
M_{2} &= \tau_{2} + \tanh r \bigl( (V_{1}+V_{2}) V_{4} - i (V_{1}-V_{2})
\tau_{1} \bigr)  \notag \\
& \quad - i \tanh^{2} r (V_{1}+V_{2})(V_{1}-V_{2}) V_{5},  \notag \\
M_{3} &= i (V_{1}-V_{3}) V_{4} + (V_{1}+V_{3}) \tau_{1}  \notag \\
& \quad + \tanh r V_{5} \frac{4V_{1}^{2} + 1 + (1-l) s^{2}}{2},  \notag \\
M_{4} &= i \tanh r (V_{1}+V_{3})(V_{1}-V_{3}),  \notag \\
M_{5} &= -i \tanh^{3} r (V_{1}+V_{2})(V_{1}-V_{2}),  \tag{C5}
\end{align}
where
\begin{align}
V_{1} &= \frac{(1-l) e^{-2x i\phi} + l}{2},  \notag \\
V_{2} &= \frac{1 + i \sqrt{1-l} s}{2},  \notag \\
V_{3} &= \frac{1 - i \sqrt{1-l} s}{2},  \notag \\
V_{4} &= t_{1} + \frac{\alpha}{\cosh r},  \notag \\
V_{5} &= t_{2} + \frac{\alpha}{\cosh r}.  \tag{C6}
\end{align}
By further utilizing Eqs.~(\ref{eq:BS2_trans}), (B1) and (C1), the
expressions for $\langle I_{D} \rangle$ and $\langle I_{D}^{2} \rangle$ can
be obtained for $k=2$ as
\begin{equation}
\langle I_{D} \rangle = i \bigl[ D_{2}(1) - D_{2}(1)^{\ast} \bigr],  \tag{C7}
\end{equation}
and
\begin{align}
\langle I_{D}^{2} \rangle &= D_{1}(0,1,0,1) + D_{1}(1,0,1,0)  \notag \\
&\quad + 2 D_{1}(1,1,1,1) - D_{2}(2) - D_{2}(2)^{\ast}.  \tag{C8}
\end{align}

\section{$C_{Q}$ for $k=2$ Case}

\label{app:D}

In this appendix, we derive $C_{Q}$ for the Kerr nonlinear phase shift with $%
k=2$ and its specific expression to obtain the QFI $F_{L_{2}}$ under photon
losses condition. To compute $C_{Q}$ using Eqs.~(\ref{eq:H1_def}), (\ref%
{eq:H2_def}), (\ref{eq:kraus_kerr}) and (\ref{eq:CQ_kerr}) according to the
method in Refs.~\cite{40} and \cite{49}, we first obtain the normal ordering
form of $(1-l)^{n_{a}} n_{a}^{q}$ by utilizing the operator identity from
Eq.~(\ref{eq:norm_order}) as follows
\begin{align}
(1-l)^{n_{a}} n_{a}^{q} &= \eta^{n_{a}} n_{a}^{q}  \notag \\
&= \frac{\partial^{q}}{\partial x^{q}} \exp\bigl[ n_{a} \ln \eta \bigr] \exp%
\bigl[ n_{a} x \bigr] \Big|_{x=0}  \notag \\
&= \colon\! \frac{\partial^{q}}{\partial x^{q}} e^{(\eta e^{x} - 1)
a^{\dagger}a} \Big|_{x=0} \!\colon,  \tag{D1}
\end{align}
where for simplicity, we set $\eta = 1-l$. Based on this equation and using
IWOP technology, the following summation expression $S_{q,p}$ can be further
calculated for the operators associated with the Hermitian operators $%
H_{1,2} $ from Eqs.~(\ref{eq:H1_def}) and (\ref{eq:H2_def}):
\begin{align}
S_{q,p} &= \sum_{j=0}^{\infty} \frac{(1-\eta)^{j}}{j!} j^{p} a^{\dagger j}
\eta^{n_{a}} n_{a}^{q} a^{j}  \notag \\
&= \sum_{j=0}^{\infty} \frac{(1-\eta)^{j}}{j!} j^{p} \colon\!
(a^{\dagger}a)^{j} \frac{\partial^{q}}{\partial x^{q}} e^{(\eta e^{x} - 1)
a^{\dagger}a} \Big|_{x=0} \!\colon  \notag \\
&= \colon\! \sum_{j=0}^{\infty} \frac{[(1-\eta) a^{\dagger}a]^{j}}{j!} \frac{%
\partial^{q+p}}{\partial x^{q} \partial y^{p}} e^{(\eta e^{x} - 1)
a^{\dagger}a + y j} \Big|_{x=y=0} \!\colon  \notag \\
&= \colon\! \frac{\partial^{q+p}}{\partial x^{q} \partial y^{p}} e^{[\eta
e^{x} + (1-\eta) e^{y} - 1] a^{\dagger}a} \Big|_{x=y=0} \!\colon  \notag \\
&= \frac{\partial^{q+p}}{\partial x^{q} \partial y^{p}} \bigl[ \eta e^{x} +
(1-\eta) e^{y} \bigr]^{n_{a}} \Big|_{x=y=0}.  \tag{D2}
\end{align}
The final step in the above equation employs the operator identity from Eq.~(%
\ref{eq:norm_order}) for conversion.

Similar to the calculation method for $C_{Q}$ in Eq.~(\ref{eq:CQ_bound}),
substituting the Kraus operator with $k=2$ (Eq.~(\ref{eq:kraus_kerr})) into
Eqs.~(\ref{eq:H1_def}) and (\ref{eq:H2_def}), and combining with Eq.~(D2)
and the following transformation relationships,
\begin{align}
e^{\lambda a^{\dagger}a} a^{j} e^{-\lambda a^{\dagger}a} &= e^{-\lambda j}
a^{j},  \notag \\
e^{\lambda (a^{\dagger}a)^{2}} a^{j} e^{-\lambda (a^{\dagger}a)^{2}} &=
e^{\lambda j^{2}} a^{j} e^{-2\lambda j a^{\dagger}a},  \tag{D3}
\end{align}
we can further compute to obtain
\begin{align}
C_{Q} &= 4 \bigl[ K_{1}^{2} \langle \Delta^{2} n_{a}^{2} \rangle_{S} - K_{2}
\langle n_{a}^{3} \rangle_{S} + K_{3} \langle n_{a}^{2} \rangle_{S}  \notag
\\
& \quad - K_{4} \langle n_{a} \rangle_{S} - K_{5} \langle n_{a}^{2}
\rangle_{S} \langle n_{a} \rangle_{S} - K_{6} \langle n_{a} \rangle_{S}^{2} %
\bigr],  \tag{D4}
\end{align}
where
\begin{align}
K_{1} &= \omega_{1} \eta^{2} - 2 \omega_{2} \eta - \mu_{2},  \notag \\
K_{2} &= 2 \eta \bigl[ 3 \omega_{1}^{2} \eta^{3} - 3 \omega_{3}^{2} \eta^{2}
- \omega_{4} \eta + \omega_{5} \bigr],  \notag \\
K_{3} &= \eta \bigl[ 11 \omega_{1}^{2} \eta^{3} - 2 \omega_{6}^{2} \eta^{2}
+ \omega_{7} \eta - 4 \omega_{1} \omega_{2} \bigr],  \notag \\
K_{4} &= \eta \omega_{1}^{2} (6 \eta^{3} - 12 \eta^{2} + 7 \eta - 1),  \notag
\\
K_{5} &= 2 (1-\eta) \eta \omega_{1} K_{1},  \notag \\
K_{6} &= (1-\eta)^{2} \eta^{2} \omega_{1}^{2},  \tag{D5}
\end{align}
and
\begin{align}
\omega_{1} &= 1 + 2 \mu_{1} - \mu_{2},  \notag \\
\omega_{2} &= \mu_{1} - \mu_{2},  \notag \\
\omega_{3} &= 1 + 2 (3 \mu_{1} - 2 \mu_{2}) + (2 \mu_{1} - \mu_{2})(4
\mu_{1} - 3 \mu_{2}),  \notag \\
\omega_{4} &= 7 \mu_{2} - 6 \mu_{1} + 24 \mu_{1} \mu_{2} - 14 \mu_{1}^{2} -
9 \mu_{2}^{2},  \notag \\
\omega_{5} &= \mu_{2} \omega_{1} - 2 \omega_{2}^{2},  \notag \\
\omega_{6} &= 9 + 40 \mu_{1} - 22 \mu_{2} + 44 \mu_{1}^{2} - 48 \mu_{1}
\mu_{2} + 13 \mu_{2}^{2},  \notag \\
\omega_{7} &= 7 + 40 \mu_{1} - 26 \mu_{2} + 52 \mu_{1}^{2} - 64 \mu_{1}
\mu_{2} + 19 \mu_{2}^{2},  \tag{D6}
\end{align}
where parameters $\mu_{1}$ and $\mu_{2}$ are optimizable to describe the
photon losses occurring before and after the phase shifter. In particular, $%
\mu_{1} = \mu_{2} = 0$ or $-1$ represents photon losses occurring before or
after the phase shifter, respectively, as depicted in Fig.~\ref{Fig7}.
Specifically, for the case $\mu_{1} = \mu_{2} = -1$, $F_{L_{2}} \leqslant
C_{Q} = 4 \langle \Delta^{2} n_{a}^{2} \rangle$, representing the ideal case
of QFI. Using Eqs.~(D4)--(D6), optimizing through $\partial C_{Q}/\partial
\mu_{1} = \partial C_{Q}/\partial \mu_{2} = 0$ to find the minimum value of $%
C_{Q}$ gives
\begin{align}
\mu_{1\text{opt}} &= \frac{G_{2} G_{5} - G_{3} G_{4}}{G_{1} G_{4} - 2\eta
G_{2}^{2}},  \tag{D7} \\
\mu_{2\text{opt}} &= \frac{G_{1} G_{5} - 2\eta G_{2} G_{3}}{G_{1} G_{4} -
2\eta G_{2}^{2}},  \tag{D8}
\end{align}
where
\begin{align}
G_{1} &= 2 \bigl[ -(1-\eta) \eta \bigl( \langle \Delta^{2} n_{a}^{2} \rangle
+ 2 \langle n_{a}^{2} \rangle \langle n_{a} \rangle - \langle n_{a}
\rangle^{2} \bigr)  \notag \\
& \quad - (6\eta^{2} - 6\eta + 1) \bigl( \langle n_{a}^{3} \rangle + \langle
n_{a} \rangle \bigr)  \notag \\
& \quad + (11\eta^{2} - 11\eta + 2) \langle n_{a}^{2} \rangle \bigr],  \notag
\\
G_{2} &= (1-\eta)^{2} \langle \Delta^{2} n_{a}^{2} \rangle + 3
(1-\eta)(2\eta-1) \langle n_{a}^{3} \rangle  \notag \\
& \quad + (11\eta^{2} - 13\eta + 3) \langle n_{a}^{2} \rangle - (6\eta^{2} -
6\eta + 1) \langle n_{a} \rangle  \notag \\
& \quad - (1-\eta)(2\eta-1) \langle n_{a}^{2} \rangle \langle n_{a} \rangle
+ \eta (1-\eta) \langle n_{a} \rangle^{2},  \tag{D9}
\end{align}
and
\begin{align}
G_{3} &= \eta^{2} \langle \Delta^{2} n_{a}^{2} \rangle - 3 \eta (2\eta-1)
\langle n_{a}^{3} \rangle  \notag \\
& \quad + (11\eta^{2} - 9\eta + 1) \langle n_{a}^{2} \rangle - (6\eta^{2} -
6\eta + 1) \langle n_{a} \rangle  \notag \\
& \quad + \eta (2\eta-1) \langle n_{a}^{2} \rangle \langle n_{a} \rangle +
\eta (1-\eta) \langle n_{a} \rangle^{2},  \notag \\
G_{4} &= -(1-\eta)^{3} \langle \Delta^{2} n_{a}^{2} \rangle - 6 \eta
(1-\eta)^{2} \langle n_{a}^{3} \rangle  \notag \\
& \quad - \eta (1-\eta)(11\eta-4) \langle n_{a}^{2} \rangle - \eta
(6\eta^{2} - 6\eta + 1) \langle n_{a} \rangle  \notag \\
& \quad + 2 \eta (1-\eta)^{2} \langle n_{a}^{2} \rangle \langle n_{a}
\rangle + \eta^{2} (1-\eta) \langle n_{a} \rangle^{2},  \notag \\
G_{5} &= \eta \bigl[ -\eta (1-\eta) \bigl( \langle \Delta^{2} n_{a}^{2}
\rangle - \langle n_{a} \rangle^{2} \bigr)  \notag \\
& \quad - (6\eta^{2} - 6\eta + 1) \bigl( \langle n_{a}^{3} \rangle + \langle
n_{a} \rangle \bigr)  \notag \\
& \quad + (11\eta^{2} - 11\eta + 2) \langle n_{a}^{2} \rangle  \notag \\
& \quad + (2\eta^{2} - 2\eta + 1) \langle n_{a}^{2} \rangle \langle n_{a}
\rangle \bigr],  \tag{D10}
\end{align}
and substituting these into Eq.~(D4), along with using Eq.~(\ref%
{eq:norm_order}) for $w=1,2,3,4$, results in $F_{L_{2}} = C_{Q\min}$.


\begin{thebibliography}{99}
\bibitem{1} Ya. M. Blanter and M. B\"uttiker, Shot noise in mesoscopic
conductors, Phys. Rep. \textbf{336}, 1 (2000).

\bibitem{2} C. M. Caves, Quantum-mechanical noise in an interferometer,
Phys. Rev. D \textbf{23}, 1693 (1981).

\bibitem{3} N. Treps, U. Andersen, B. Buchler, P. K. Lam, A. Ma\^{\i}tre, H.
A. Bachor, and C. Fabre, Surpassing the Standard Quantum Limit for Optical
Imaging Using Nonclassical Multimode Light, Phys. Rev. Lett. \textbf{88},
203601 (2002).

\bibitem{4} D. Bouwmeester, J. W. Pan, K. Mattle, M. Eibl, H. Weinfurter,
and A. Zeilinger, Experimental quantum teleportation, Nature \textbf{390},
575 (1997).

\bibitem{5} A. Ac\'{\i}n, N. Gisin, and L. Masanes, From Bell's Theorem to
Secure Quantum Key Distribution, Phys. Rev. Lett. \textbf{97}, 120405 (2006).

\bibitem{6} B. Kraus, C. O. Ahonen, M. M\"ott\"onen, and J. L. O'Brien,
Entanglement-enhanced quantum key distribution, Phys. Rev. A \textbf{78},
032314 (2008).

\bibitem{7} L. Y. Hu, M. Al-amri, Z. Y. Liao, and M. S. Zubairy,
Continuous-variable quantum key distribution with non-Gaussian operations,
Phys. Rev. A \textbf{102}, 012608 (2020).

\bibitem{8} Z. Zhang, S. Mouradian, C. Wong, F. N. C. Wong, and J. H.
Shapiro, Entanglement-Enhanced Sensing in a Lossy and Noisy Environment,
Phys. Rev. Lett. \textbf{114}, 110506 (2015).

\bibitem{9} V. Giovannetti, S. Lloyd, and L. Maccone, Quantum-Enhanced
Measurements: Beating the Standard Quantum Limit, Science \textbf{306}, 1330
(2004).

\bibitem{10} H. Lee, P. Kok, and J. P. Dowling, A quantum Rosetta stone for
interferometry, J. Mod. Opt. \textbf{49}, 2325 (2002).

\bibitem{11} J. P. Dowling, Quantum optical metrology---the lowdown on
high-N00N states, Contemp. Phys. \textbf{49}, 125 (2008).

\bibitem{12} H. Cable and G. A. Durkin, Parameter Estimation with Entangled
Photons Produced by Parametric Down-Conversion, Phys. Rev. Lett. \textbf{105}%
, 013603 (2010).

\bibitem{13} K. Jiang, C. J. Brignac, Y. Weng, M. B. Kim, H. Lee, and J. P.
Dowling, Strategies for choosing path-entangled number states for optimal
robust quantum-optical metrology in the presence of loss, Phys. Rev. A
\textbf{86}, 013826 (2012).

\bibitem{14} M. W. Mitchell, J. S. Lundeen, and A. M. Steinberg,
Super-resolving phase measurements with a multiphoton entangled state,
Nature \textbf{429}, 161 (2004).

\bibitem{15} I. Afek, O. Ambar, and Y. Silberberg, High-NOON States by
Mixing Quantum and Classical Light, Science \textbf{328}, 879 (2010).

\bibitem{16} J. C. F. Matthews, A. Politi, D. Bonneau, and J. L. O'Brien,
Heralding Two-Photon and Four-Photon Path Entanglement on a Chip, Phys. Rev.
Lett. \textbf{107}, 163602 (2011).

\bibitem{17} A. Zavatta, S. Viciani, and M. Bellini, Quantum-to-Classical
Transition with Single-Photon-Added Coherent States of Light, Science
\textbf{306}, 660 (2004).

\bibitem{18} F. Jia, W. Ye, Q. Wang, L. Y. Hu, and H. Y. Fan, Comparison of
nonclassical properties resulting from non-Gaussian operations, Laser Phys.
Lett. \textbf{16}, 015201 (2019).

\bibitem{19} H. Zhang, W. Ye, C. P. Wei, Y. Xia, S. K. Chang, Z. Y. Liao,
and L. Y. Hu, Improved phase sensitivity in quantum optical interferometer
based on multi-photon catalytic two-mode squeezed vacuum states, Phys. Rev.
A \textbf{103}, 013705 (2021).

\bibitem{20} H. Zhang, W. Ye, C. P. Wei, C. J. Liu, Z. Y. Liao, and L. Y.
Hu, Improving phase estimation using number-conserving operations, Phys.
Rev. A \textbf{103}, 052602 (2021).

\bibitem{21} Y. K. Xu, S. K. Chang, C. J. Liu, L. Y. Hu, and S. Q. Liu,
Phase estimation of an SU(1,1) interferometer with a coherent superposition
squeezed vacuum in a realistic case, Opt. Express \textbf{30}, 38178 (2022).

\bibitem{22} Y. K. Xu, T. Zhao, Q. Q. Kang, C. J. Liu, L. Y. Hu, and S. Q.
Liu, Phase sensitivity of an SU(1,1) interferometer in photon-loss via
photon operations, Opt. Express \textbf{31}, 8414 (2023).

\bibitem{23} L. L. Guo, Y. F. Yu, and Z. M. Zhang, Improving the phase
sensitivity of an SU(1,1) interferometer with photon-added squeezed vacuum
light, Opt. Express \textbf{26}, 29099 (2018).

\bibitem{24} S. Wang, X. X. Xu, Y. J. Xu, and L. J. Zhang, Quantum
interferometry via a coherent state mixed with a photon-added squeezed
vacuum state, Opt. Commun. \textbf{444}, 102--110 (2019).

\bibitem{25} Y. Ouyang, S. Wang, and L. J. Zhang, Quantum optical
interferometry via the photon-added two-mode squeezed vacuum states, J. Opt.
Soc. Am. B \textbf{33}, 1373 (2016).

\bibitem{26} R. Birrittella and C. C. Gerry, Quantum optical interferometry
via the mixing of coherent and photon-subtracted squeezed vacuum states of
light, J. Opt. Soc. Am. B \textbf{31}, 586 (2014).

\bibitem{27} Q. Q. Kang, Z. K. Zhao, T. Zhao, C. J. Liu, and L. Y. Hu, Phase
estimation via number-conserving operation inside the SU(1,1)
interferometer, Phys. Rev. A \textbf{110}, 022432 (2024).

\bibitem{28} C. Kumar, Rishabh, and S. Arora, Realistic
non-Gaussian-operation scheme in parity-detection-based Mach-Zehnder quantum
interferometry, Phys. Rev. A \textbf{105}, 052437 (2022).

\bibitem{29} Y. J. Chen, J. W. Gao, J. X. Han, Z. H. Yuan, R. Q. Li, Y. Y.
Jiang, and J. Song, Orbital-angular-momentum-enhanced phase estimation using
non-Gaussian states with photon loss, Phys. Rev. A \textbf{108}, 022613
(2023).

\bibitem{30} Z. K. Zhao, Q. Q. Kang, H. Zhang, T. Zhao, C. J. Liu, and L. Y.
Hu, Phase estimation via coherent and photon-catalyzed squeezed vacuum
states, Opt. Express \textbf{32}, 28267 (2024).

\bibitem{31} Q. Q. Kang, Z. K. Zhao, Y. K. Xu, T. Zhao, C. J. Liu, and L. Y.
Hu, Phase estimation based on multi-photon subtraction operation inside the
SU(1,1) interferometer, Phys. Scripta \textbf{99}, 085111 (2024).

\bibitem{32} C. Kumar, Rishabh, M. Sharma, and S. Arora,
Parity-detection-based Mach-Zehnder interferometry with coherent and
non-Gaussian squeezed vacuum states as inputs, Phys. Rev. A \textbf{108},
012605 (2023).

\bibitem{33} S. Boixo, A. Datta, S. T. Flammia, A. Shaji, E. Bagan, and C.
M. Caves, Quantum Metrology: Dynamics versus Entanglement, Phys. Rev. A
\textbf{77}, 012317 (2008).

\bibitem{34} J. Joo, K. Park, H. Jeong, W. J. Munro, K. Nemoto, and T. P.
Spiller, Quantum metrology for nonlinear phase shifts with entangled
coherent states, Phys. Rev. A \textbf{86}, 043828 (2012).

\bibitem{35} C. P. Wei and Z. M. Zhang, Improving the phase sensitivity of a
Mach--Zehnder interferometer via a nonlinear phase shifter, J. Mod. Opt.
\textbf{64}, 743 (2017).

\bibitem{36} J. D. Zhang and S. Wang, Nonlinear phase estimation based on
nonlinear interferometers with coherent and squeezed vacuum light, Phys.
Lett. A \textbf{502}, 129400 (2024).

\bibitem{37} G. F. Jiao, K. Y. Zhang, L. Q. Chen, W. P. Zhang, and C. H.
Yuan, Nonlinear phase estimation enhanced by an actively correlated
Mach-Zehnder interferometer, Phys. Rev. A \textbf{102}, 033520 (2020).

\bibitem{38} G. F. Jiao, Q. Wang, Z. F. Yu, L. Q. Chen, W. P. Zhang, and C.
H. Yuan, Effects of losses on the sensitivity of an actively correlated
Mach-Zehnder interferometer, Phys. Rev. A \textbf{104}, 013725 (2021).

\bibitem{39} S. K. Chang, C. P. Wei, H. Zhang, Y. Xia, W. Ye, and L. Y. Hu,
Enhanced phase sensitivity with a nonconventional interferometer and
nonlinear phase shifter, Phys. Lett. A \textbf{384}, 126755 (2020).

\bibitem{40} S. K. Chang, W. Ye, H. Zhang, L. Y. Hu, J. H. Huang, and S. Q.
Liu, Improvement of phase sensitivity in an SU(1,1) interferometer via a
phase shift induced by a Kerr medium, Phys. Rev. A \textbf{105}, 033704
(2022).

\bibitem{41} J. D. Zhang, Z. J. Zhang, L. Z. Cen, J. Y. Hu, and Y. Zhao,
Nonlinear phase estimation: Parity measurement approaches the quantum
Cram\'er-Rao bound for coherent states, Phys. Rev. A \textbf{99}, 022106
(2019).

\bibitem{42} K. P. Seshadreesan, P. M. Anisimov, H. Lee, and J. P. Dowling,
Parity detection achieves the Heisenberg limit in interferometry with
coherent mixed with squeezed vacuum light, New J. Phys. \textbf{13}, 083026
(2011).

\bibitem{43} B. Yurke, S. L. McCall, and J. R. Klauder, SU(2) and SU(1,1)
interferometers, Phys. Rev. A \textbf{33}, 4033 (1986).

\bibitem{44} S. L. Braunstein and C. M. Caves, Statistical distance and the
geometry of quantum states, Phys. Rev. Lett. \textbf{72}, 3439 (1994).

\bibitem{45} S. L. Luo, Wigner-Yanase skew information and uncertainty
relations, Phys. Rev. Lett. \textbf{91}, 180403 (2003).

\bibitem{46} Z. Y. Ou, Complementarity and fundamental limit in precision
phase measurement, Phys. Rev. Lett. \textbf{77}, 2352 (1996).

\bibitem{47} C. W. Helstrom, \textit{Quantum detection and estimation theory}
(Academic, New York, 1976).

\bibitem{48} J. Liu, X. X. Jing, W. Zhong, and X. G. Wang, Quantum Fisher
information for density matrices with arbitrary ranks, Commun. Theor. Phys.
\textbf{61}, 45 (2014).

\bibitem{49} B. M. Escher, R. L. de Matos Filho, and L. Davidovich, General
framework for estimating the ultimate precision limit in noisy
quantum-enhanced metrology, Nat. Phys. \textbf{7}, 406 (2011).
\end{thebibliography}
\end{document}